\documentclass[preprint2]{proto}
\usepackage{times,lscape}

\newcommand{\refs}{\par\noindent\hangindent=1pc\hangafter=1}
\begin{document}
\newcommand\go{\rm $G_0$}
\newcommand\msunyr{\rm M_{\odot}\,yr^{-1}}
\newcommand\msun{\rm M_{\odot}}
\newcommand\lsun{\rm L_{\odot}}
\newcommand\rsun{\rm R_{\odot}}
\newcommand\mdot{ \dot{M}}
\newcommand\etal{et al. }
\newcommand\be {\begin{equation}}
\newcommand\en{\end{equation}}
\newcommand\cm{\rm cm}
\newcommand\din{\rm dyn}
\newcommand\gr{\rm g}
\newcommand\seg{\rm s}
\newcommand\AU{\rm AU}
\newcommand\Myr{\rm Myr}
\newcommand\K{\rm K}
\newcommand\yrs{\rm yrs}
\newcommand\av{\rm $A_V$}

\title{\textbf{\LARGE The Chemical Evolution of Protoplanetary Disks}}

\author {\textbf{\large Edwin A. Bergin}}
\affil{\small\em University of Michigan}
\author {\textbf{\large Yuri Aikawa}}
\affil{\small\em Kobe University}
\author {\textbf{\large Geoffrey A. Blake}}
\affil{\small\em California Institute of Technology}
\author {\textbf{\large Ewine F. van Dishoeck}}
\affil{\small\em Leiden Observatory}

\begin{abstract}
\begin{list}{ } {\rightmargin 1in}
\baselineskip = 11pt
\parindent=1pc
{\small
In this review we re-evaluate our observational
and theoretical understanding of
the chemical evolution of protoplanetary disks.
We discuss how
improved observational capabilities have enabled the detection of numerous
molecules exposing an active disk chemistry
that appears to be in disequilibrium.
We outline the primary facets of static and dynamical
theoretical chemical models.  Such models have
demonstrated that the observed disk chemistry arises from warm surface
layers that are irradiated by X-ray and FUV emission from the central
accreting star.  Key emphasis is placed on reviewing areas where disk chemistry
and physics are linked: including the deuterium chemistry,
gas temperature structure,
disk viscous evolution (mixing), ionization fraction, and the beginnings of planet formation.
 \\~\\~\\~}
\end{list}
\end{abstract}

\section{\textbf{INTRODUCTION}}

For decades models of our own Solar nebular chemical and physical evolution
have been constrained by the chemical record gathered from meteorites, planetary
atmospheres, and cometary comae.
Such studies have
provided important clues to the formation of the sun and planets, but large
questions remain regarding the structure of the solar nebula, the
exact timescale of planetary formation, and the chemical evolution of nebular
gas and dust.  Today we are on the verge of a different approach to nebular
chemical studies, one where the record gained by solar system studies is combined
with observations of numerous molecular lines
in a multitude of extra-solar  protoplanetary disk systems
tracking various evolutionary stages.

Our observational understanding  of extra-solar protoplanetary disk
systems is still in its infancy as the
current capabilities of millimeter-wave observatories
are limited by sensitivity and also by the small angular size of
circumstellar disks, even in the closest star-forming regions ($< 3-4''$).
Nonetheless, numerous molecules have been detected in protoplanetary disks,
exposing an active chemistry ({\it Dutrey et al.}, 1997; {\it Kastner et al.}, 1997; {\it Qi et
al.}, 2003;
{\it Thi et al.}, 2004).
Since the last {\em Protostars \& Planets} review ({\it Prinn}, 1993) these
observations have  led to a paradigm shift in our  understanding
of disk chemistry.
For many years focus was placed on thermochemical
models as predictors of the gaseous composition, and these models have
relevance in the high pressure, $\gtrsim 10^{-6}$ bar, (inner) regions of the nebula (e.g. {\it
Fegley}, 1999).
However, for most of the disk mass, the observed chemistry appears to be
in disequilibrium and quite similar to that
seen in dense regions of the interstellar medium (ISM) that are directly exposed
to radiation ({\it Aikawa and Herbst, 1999}; {\it Willacy and Langer}, 2000; {\it Aikawa et al.},
2002).

In this review we focus on gains in our understanding of the chemistry
that precedes and is contemporaneous with the formation of planets from a
perspective guided by observations of other stellar systems whose
masses are similar to the Sun.
We examine both the observational and theoretical aspects of this
emerging field, with an
emphasis on areas where the chemistry directly relates to disk physics.
Portions of this review overlap with other chapters, such as the
observational summary of molecular disks,
inner disk gas, and disk physical structure 
(see the Chapters by Dutrey et al., 
Najita et al., and Dullemond et al., respectively).
For this purpose we
focus our review on the physics and chemistry of the
outer disk ($r >$ 10 AU) in systems with ages of 0.3-10 Myr. In support of theory we
also present an observational perspective extending from the infrared (IR) to the
submillimeter (sub-mm) to motivate the theoretical background and supplement other discussions.

\section{\textbf{GENERAL THEORETICAL PICTURE}}

\subsection{Basic Physical and Chemical Structure of Disk}


Chemical abundances are determined by physical conditions such as
density,
temperature, and the incident 

\begin{landscape}
\begin{figure}
\centering
\includegraphics[width=21cm]{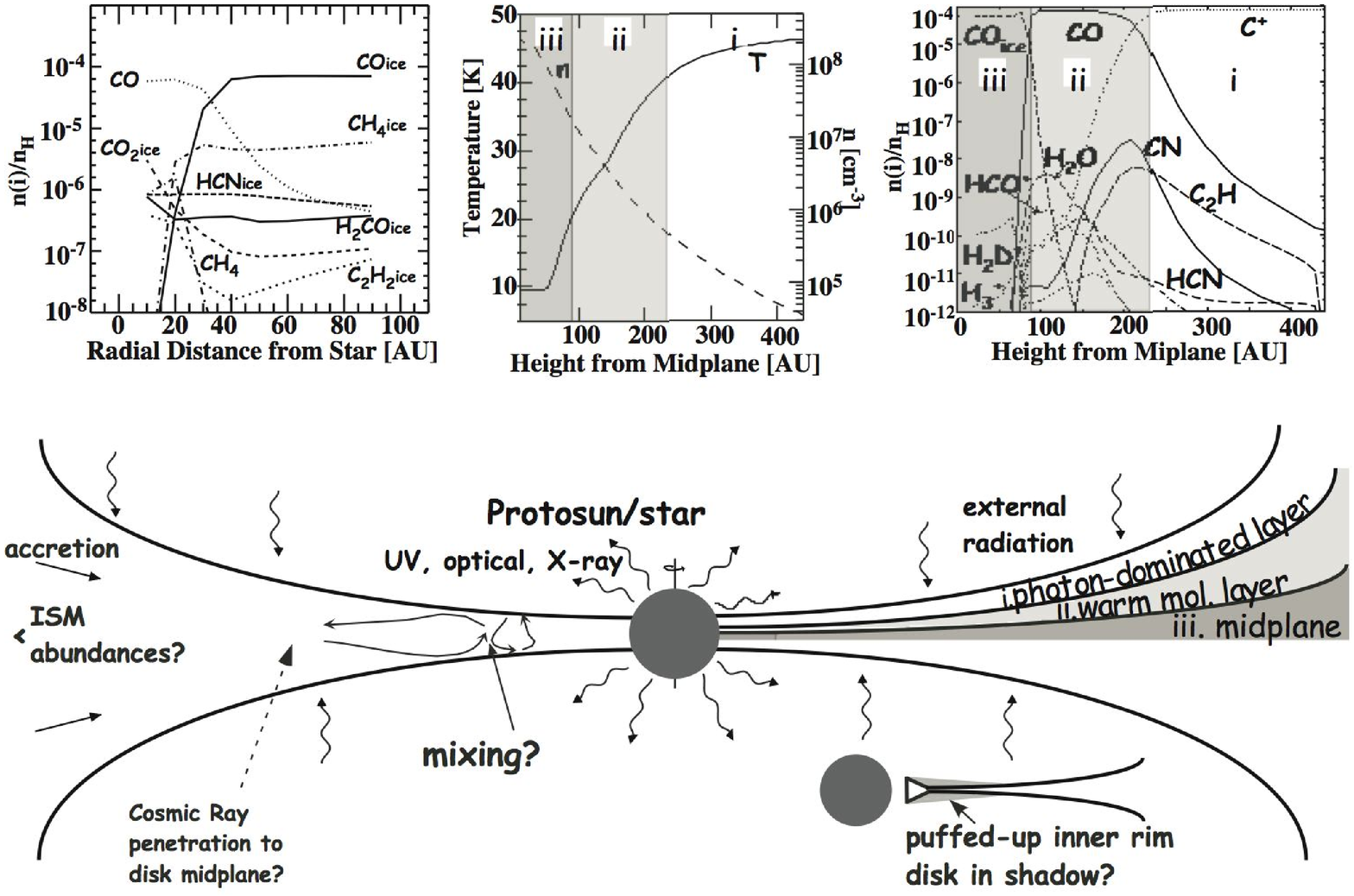}
 \caption{\small Chemical structure of protoplanetary disks. Vertically the disk is
schematically divided into three zones: a photon-dominated layer, a warm molecular
layer, and a midplane freeze-out layer. The CO freeze-out layer disappears at
$r\lesssim 30-60$ AU as the mid-plane temperature increases inwards. Various non-thermal inputs,
cosmic ray, UV, and X-ray drive chemical reactions. Viscous accretion and
turbulence will transport the disk material both vertically and radially.
The upper panels show the radial and vertical distribution of molecular
abundances from a typical disk model at the midplane 
({\it Aikawa et al.}, 1999) and $r\sim 300$ AU ({\it van Zadelhoff et al.},
2003).  A sample of the hydrogen density and {\em dust} temperature at the
same distance ({\it D'Alessio et al.}, 1999) is also provided.  
In upper layers ($\gtrsim 150$ AU)
the {\em gas} temperature will exceed the dust temperature by $\gtrsim 25$ K
({\it Jonkheid et al.} 2004).
}
\label{overview}
\end{figure}
\end{landscape}

\noindent 
radiation field. Recent years have 
seen significant progress in characterizing
disk physical structure, which aids in
understanding disk chemical processes.
Isolated disks can be quite extended with $r_{out} \sim$ one hundred to a 
few hundred AU 
({\it Simon, Dutrey, and Guilloteau} 2000), much larger than expected from comparison
with the Minimum Mass Solar Nebula (MMSN: {\it Hayashi}, 1981).
Although, it should be stated that we have an observational bias towards detecting larger
disks due to our observational limitations.
The radial distribution of column density and midplane temperature have been estimated by
observing thermal emission of dust; they are fitted by a power-law
$\Sigma(r)\propto r^{-p}$ and $T(r)\propto r^{-q}$, with $p=0-1$ and
$q=0.5-0.75$. The temperature at 1~AU is $\sim 100-200$ K, while the surface
density at 100~AU is $0.1-10$ g cm$^{-2}$  (e.g. {\it Beckwith et al.}, 1990;
{\it Kitamura et al.}, 2002).

The vertical structure
is estimated
by calculating the hydrostatic equilibrium for the density and radiation
transfer for the dust temperature (see the Chapter by
{\it Dullemond et al.}). Beyond several AU
the disk is mainly heated by irradiation from the central star. The stellar
radiation is absorbed by grains at the disk surface, which then emit thermal radiation
to heat the disk interior ({\it Calvet et al.}, 1992;
{\it Chiang and Goldreich}, 1997; {\it D'Alessio et al.}, 1998). Hence the temperature
decreases towards the midplane, as seen in Fig.~1.
For small radii ($r <$ few AU), however,  heating by
mass accretion is not negligible and the midplane can be warmer than the
disk surface.
At $r=1$ AU, for example, the midplane temperature can be as
high as 1000 K, if the accretion rate is large ({\it D'Alessio et al.}, 1999; {\it Nomura},
2002). The density distribution is basically Gaussian, $\exp {-(Z/H)^2}$,
with some deviation due to vertical
temperature variations (see Fig.~1).
As a whole the disk has a flared-up structure,
with a geometrical thickness that increases with radius ({\it Kenyon and Hartmann}, 1987).


Based on such physical models, the current picture of the general disk chemical structure
is schematically shown in Fig.~\ref{overview}. At $r\gtrsim 100$ AU, the disk
can be divided into three layers: the photon dominated region (PDR), the warm
molecular layer, and the midplane freeze-out layer.
The disk is irradiated by UV radiation from the central
star and interstellar radiation field that ionize and dissociate molecules
and atoms in the surface layer.
In the midplane the temperature is mostly lower than the freeze-out temperature
of CO ($\sim 20$ K), one of the most abundant and volatile molecules
in the ISM.
Since the timescale of adsorption onto grains is short at
high density ($\sim 10 (10^9 {\rm cm}^{-3}/n_{\rm H})$ yr),
heavy-element species are significantly depleted
onto grains. At intermediate heights, the temperature is
several 10s of K, and the density is sufficiently high ($\gtrsim 10^6$
cm$^{-3}$) to ensure the existence of molecules even if the UV radiation is
not completely attenuated by the upper layer ({\it Aikawa and Herbst}, 1999;
{\it Willacy and Langer}, 2000; {\it Aikawa et al.}, 2002).
Here water is still frozen onto grains, trapping much of the oxygen
in the solid state.  Thus, the warm CO-rich gas layers will have C/O $\sim$ 1,
leading to a rich and extensive carbon-based chemistry.

\begin{figure*}
 \epsscale{1.5}
 \plotone{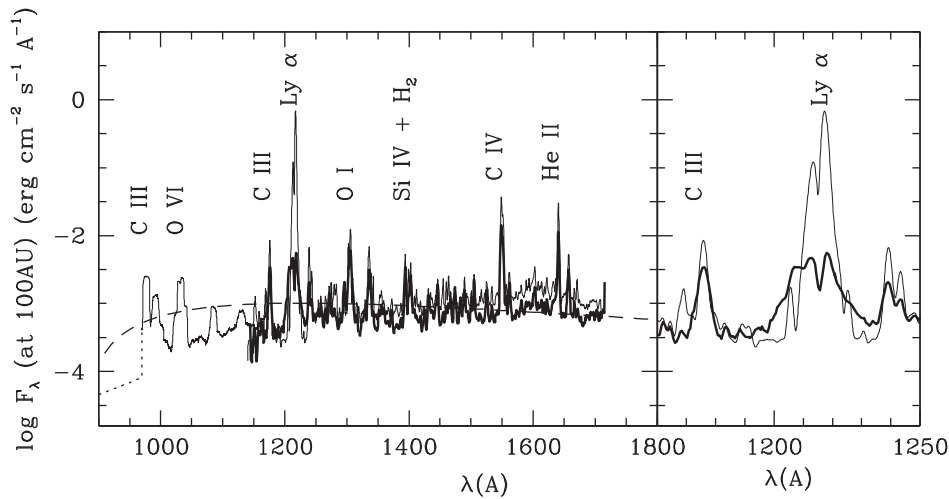}
 \caption{\small UV spectra of T Tauri stars. Heavy solid lines and light
solid lines represent the spectra of BP Tau and TW Hya, respectively.
The spectrum of TW Hya is scaled by 3.5 to match the BP Tau continuum level.
The long dashed line represents the interstellar radiation field of {\it Draine} (1978) scaled by a factor of 540.
The region around the Ly$\alpha$ line is enlarged in the right panel.
Taken from {\it Bergin et al.} (2003).
 \label{uv}}
 \end{figure*}

These models provide a good match
to observed abundances. {\it Dutrey et al.}, (1997) found that in the
DM Tau disk,
molecular abundances are generally
lower than in dense clouds, but the CN/HCN ratio is higher
(see also {\it Thi et al.,} 2004). The low molecular
abundances are caused by depletion in the midplane, and the high
CN/HCN ratio originates in the surface PDR (cf. Fig.~\ref{overview}),
as seen in PDRs in the ISM
({\it Rodr\'iguez-Franco et al.}, 1998).

At $r\lesssim$ 100 AU, the midplane temperature is high
enough to sublimate various ice materials that formed originally in
the outer disk radius and/or parental cloud core (e.g. {\it Markwick et al.}, 2002).
This sublimation will be species dependent with the ``snow line'' for a given
species appearing at different radii.  For example, in the solar nebula the water
ice snow line appeared near 3--5 AU, while the CO snow line would appear at
greater distances where the midplane dust temperatures drop below $\sim$20 K.
Within these species-selective gaseous zones,
sublimated molecules will be destroyed and transformed to other molecules by
gas-phase reactions. In this fashion, the chemistry is similar to
the so-called ``hot core'' chemistry,
which appears in star-forming cores surrounding protostars
(e.g. {\it Hatchell et al.}, 1998; see the Chapter by {\it Ceccarelli et al.}).
For example, sublimated CH$_4$ is transformed to larger and less volatile
carbon-chain species, which can then accumulate onto grains ({\it Aikawa et al.},
1999).

\subsection{Key Ingredients: UV, X-ray, and Cosmic-ray}

Thermochemistry and its consequences are
described by {\it Fegley et al.} (1989) and {\it Prinn et al.} (1993).
In recent years non-thermal events such as cosmic rays, UV and X-rays have
been included in the disk chemistry models and shown to play important roles.

In molecular clouds, chemical reactions are driven by cosmic-ray ionization
(e.g. {\it Herbst and Klemperer}, 1973). Since cosmic-ray ionization has an attenuation
length of 96 g cm$^{-2}$ ({\it Umebayashi and Nakano}, 1981), in the disk it can be important,
for $r \gtrsim$ several AU, in
driving ion-molecule reactions
and producing radicals which undergo neutral-neutral reactions
({\it Aikawa et al.}, 1997).
Although cosmic rays may be scattered by the magnetic fields within and around
the protostar-disk system, detection of ions (e.g. HCO$^+$ and
H$_2$D$^+$) and comparison with theoretical models indicate
that some ionization mechanisms, perhaps cosmic-rays, 
is available at least for $r \gtrsim$ 100 AU
({\it Semenov et. al.} 2004; {\it Ceccarelli and Dominik} 2005) (see also \S~\ref{sec:ionfrac}).

T Tauri stars have excess UV flux that is much higher than expected from
their effective temperature of $\sim 3000$~K (e.g. {\it Herbig and Goodrich}, 1986).
It is considered to originate in the accretion shock on the stellar surface
({\it Calvet and Gullbring}, 1998), with potential contributions from an active
chromosphere ({\it Alexander, Clarke, and Pringle}, 2005).
For low-mass T Tauri stars, the strength of the UV radiation field is often parameterized in
terms of the local interstellar radiation field ({\it Habing}, 1968; $G_0 = 1$), and have
values of $G_0 \sim 300-1000$ at 100 AU ({\it Bergin et al.}, 2004).
This radiation impinges on the flared disk surface with a shallow angle of
incidence.
Stellar and interstellar UV photons dissociate and ionize molecules
and atoms in the flared disk surfaces and
detailed two-dimensional radiative transfer models are required to
quantitatively predict molecular abundances. {\it van Zadelhoff et al.} (2003)
showed scattering of stellar UV at the disk surface significantly enhances the
abundance of radical species in deeper layers, and examined the resulting
chemical evolution for different wavelength dependencies of the stellar
UV radiation field.
{\it Bergin et al.} (2003) pointed out the importance of Ly~$\alpha$
emission, which stands out in the UV spectrum of T Tauri stars
(Fig.~\ref{uv}) and is absent from the interstellar field. 
In the one case where the line is unaffected by
interstellar absorption, TW Hya, Ly~$\alpha$ radiation carries $\sim 85$\% of
the FUV flux ({\it Herczeg et al.} 2004).
Since photodissociation cross sections are a function of
wavelength, species that absorb Lyman $\alpha$, such as HCN
and H$_2$O, will be selectively dissociated, while others, such as CO and
H$_2$, are
unaffected ({\it van Dishoeck, Jonkheid, and van Hemert}, 2006).  

Most stars do not form in isolation, rather 70-90\% stars are born in GMC's
(which contain most of the Galactic molecular mass)
and are found in embedded stellar clusters ({\it Lada and Lada} 2003).
In this light there exists growing evidence that the Sun formed in a
cluster
in the vicinity of a massive star (see the Chapter by {\it Wadhwa et al.};
{\it Hester and Desch} 2006).  In this case
the UV radiation field can be much higher, depending on the spectral
type
of the OB star, the proximity of the low mass star to the source of
energetic radiation, and on the dissipation timescale of the surrounding
dust and gas (which can shield forming low mass disks from
radiation).
After gas/dust dissipation the external radiation has greater
penetrating power, because it can impinge on the disk with a
greater angle of incidence.  The
primary effects of external radiation, if it dominates 
the stellar contribution, will be to magnify the chemistry (e.g. 
CO driven photo-chemistry) produced by the
stellar radiation and increase the size of the warm molecular layer.

Strong X-ray emission is observed toward T Tauri stars (e.g. {\it
  Kastner et al.}, 2005). It may originate in the magnetic
  reconnections either in the stellar magnetosphere, at the star-disk
  interface, or above the circumstellar disk ({\it Feigelson and
  Montmerle}, 1999 and references therein). X-rays affect the
  chemistry in several ways ({\it Maloney et al.}, 1996; {\it
  St{\"a}uber et al.}, 2005).  (1) They ionize atoms and molecules to
  produce high-energy photoelectrons that further ionize the gas. On
  the disk surface X-ray ionization produces a higher ionization rate
  than cosmic-rays, and can even be the dominant ionization source if
  the cosmic-rays are scattered by the magnetic field ({\it Glassgold
  et al.}, 1997; {\it Igea and Glassgold}, 1999).  (2) High-energy
  photoelectrons heat the gas. For example, at $r=1$ AU, the gas
  temperature in the upper most layer can be as high as 5000 K due to
  the X-ray heating together with mechanical heating such as turbulent
  dissipation ({\it Glassgold et al.}, 2004; see the Chapter by {\it
  Najita et al.}).  (3) Collision of the high-energy electrons with
  hydrogen atoms and molecules results in excitation of these species
  and then the emission of UV photons within the disk ({\it Maloney et
  al.}, 1996; {\it Herczeg et al.}, 2004). Recently {\it Bergin et
  al.}, (2004) found H$_2$ FUV continuum emission caused by this
  mechanism. The high ionization rate and induced photodissociation of
  CO enhance the abundances of organic species such as CN, HCN. and
  HCO$^+$ ({\it Aikawa and Herbst}, 1999; 2001).

Non-thermal particles and radiation can also drive the
desorption of molecular species from the grain surface. Because of the high
densities, gaseous species collide with grains on short timescales.
In the low-temperature region in the outer disk
the colliding molecules are adsorbed onto grains, and thermal desorption is
inefficient except for very volatile species such as CO, N$_2$, and CH$_4$,
depending on the temperature. Various gaseous molecules are still observable
since they are formed from CO and N$_2$ via gas-phase reactions, that
compensates
for adsorption (e.g. {\it Aikawa et al.} 2002). However, observations may
indicate the need more efficient (thermal or non-thermal) desorption, 
especially for species
mainly formed by grain-surface reactions. For example, {\it Dartois et al.} (2003)
find evidence for gaseous CO in layers with a temperature
below the CO sublimation temperature.
Cosmic-rays and X-rays can temporally
`spot heat' the grains to enhance desorption rates (e.g. {\it L\'eger et al.}, 1985;
{\it Hasegawa and Herbst}, 1993). UV radiation can also desorb molecules, possibly by
producing radicals within the ice mantle, which react with other radicals on
the grain surface to release excess energies ({\it Westley  et al.}, 1995).
It should be noted, however, that
the non-thermal desorption rates are uncertain and depend on various
parameters such as structure of the grain particle ({\it Najita et al.}, 2001),
UV flux, number density of radical species in grain mantle ({\it Shen  et al.},
2004), and detailed desorption process ({\it Bringa and Johnson}, 2004).
Mixing could also play a role in moving material from warmer layers to
colder ones (e.g. \S 4.4).

\section{Observations}

Observational studies of the chemistry in extrasolar protoplanetary
disks started only in the last decade thanks to improved sensitivity
and spatial resolution at millimeter and IR wavelengths.  At
millimeter wavelengths, molecules other than CO have now been detected
and imaged in a handful of disks with single-dish telescopes and
interferometers. This technique has the advantage that molecules with
very low abundances (down to $10^{-11}$ with respect to H$_2$) can be
detected through their pure rotational transitions, and that the
spatial distribution in the disk can be determined.  With a spectral
resolving power $R\approx \lambda/\Delta \lambda > 10^6$, the line
profiles are fully resolved and kinematic information can be derived.
Infrared spectroscopy has the main advantage that not only gas but
also solid material can be probed through their vibrational
transitions, including ices and silicates.  Also, gas-phase molecules
without dipole moments, including H$_2$, CH$_4$, C$_2$H$_2$ and
CO$_2$, can only be observed at IR wavelengths.  Finally,
Polycyclic Aromatic Hydrocarbons (PAHs) have unique IR
features. For space-based instruments, the resolving power is usually
low, typically $R\approx 300-3000$, making it difficult to observe and
resolve gas-phase lines.

\subsection{Infrared observations}

\subsubsection{Silicates, ices and PAHs}

The {\it Infrared Space Observatory} (ISO) opened up mid-IR
spectroscopy of disks over the full 2--200 $\mu$m range unhindered by
the Earth's atmosphere, revealing a wealth of features (see {\it van
Dishoeck}, 2004 for a review). Because of limited sensitivity, ISO
could only probe the chemistry in disks around intermediate-mass
Herbig Ae/Be stars. The {\it Spitzer Space Telescope} has the
sensitivity to take 5--40 $\mu$m spectra of solar-mass T Tauri stars,
while large 8--10m optical telescopes can obtain higher spectral and
spatial resolution data in atmospheric windows, most notably at 3--4,
4.6--5 and 8--13 $\mu$m. The features are usually in emission, except
if the disk is viewed nearly edge-on when the bands occur in
absorption against the continuum of the warm dust in the inner disk.

\begin{figure*}
\centering
\includegraphics[width=14.0cm]{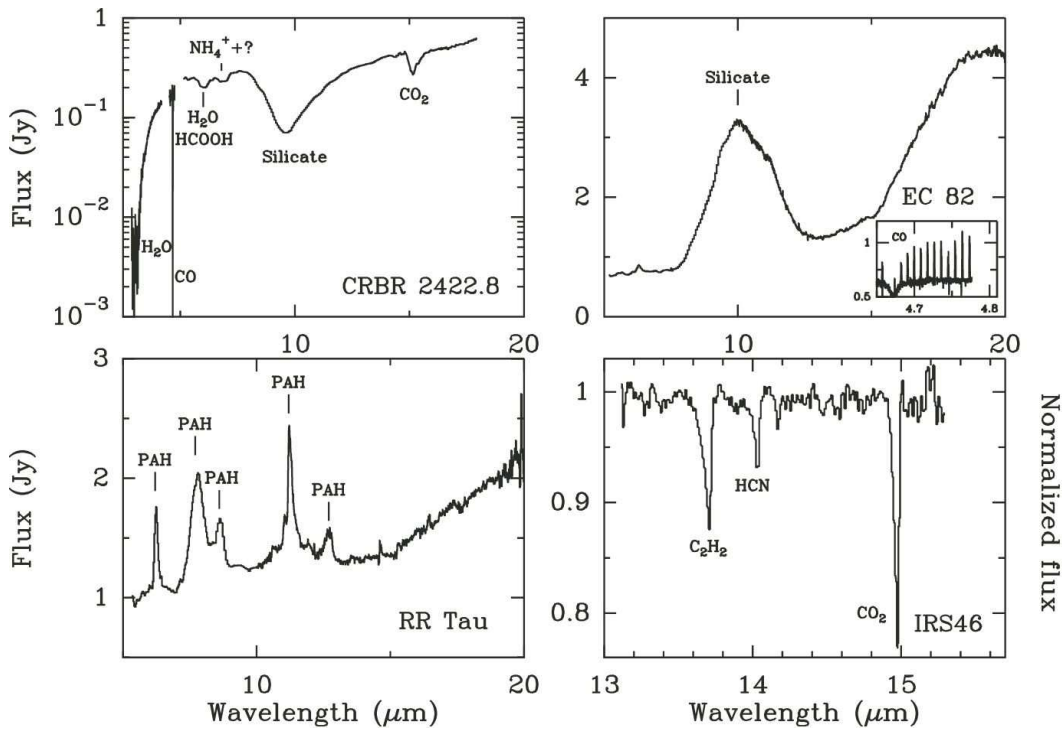}
\caption{
{\em Top left:} ice features toward edge-on
disk CRBR 2422.8 -3423. Some absorptions arise 
in the cold foreground core 
({\it Pontoppidan et al.}, 2005). {\em Top right:} silicate emission at 10
and 20 $\mu$m toward EC 82 ({\it Kessler-Silacci et al.}, 2006); inset,
gas-phase CO $v$=1-0 emission
({\it Blake and Boogert}, priv. comm.). {\em Bottom right:} PAH features
toward RR Tau ({\it Geers et al.}, 2006, in prep.);  {\em Bottom left:} gaseous C$_2$H$_2$,
HCN and CO$_2$ toward IRS 46 in Oph ({\it Lahuis et al.}, 2006);
}
\end{figure*}

The amorphous broad silicate features at $\sim$10 and 20 $\mu$m are
the most prominent emission bands in disk spectra (e.g., {\it Przygodda et
al.}\ 2003, {\it Kessler-Silacci et al.}\ 2005; 2006) (see Fig.~3). They arise in
the super-heated layers of the inner disk at $<$1--10 AU and are
not representative of the outer disk. For at
least half of the sources narrower features are seen as well, which
can be ascribed to crystalline silicates such as forsterite,
Mg$_2$SiO$_4$ (e.g., {\it Malfait et al.},  1998, {\it Forrest et al.},
2004). Silicates are discussed extensively in the Chapter by {\it Natta et 
al.}, the main point for this chapter is that the shape
and strength of the silicate features, together with the overall
spectral energy distribution, can be used to determine whether grain
growth and settling has occurred 
(e.g., {\it van Boekel et
al.}, 2005). This, in turn, affects the chemistry and heating in the
disk (see \S 4.1).
Also, the presence of crystalline silicates may be
an indication of significant radial and vertical mixing (see \S 4.4).

Ices can only be present in the cold, outer parts of the disk where
the temperature drops below 100 K. Thus, the strongest ice emission
bands typically occur at far-IR wavelengths. Crystalline water
ice has been seen in a few disks through librational features at 44
and 63 $\mu$m (e.g., {\it Malfait et al.}, 1998, {\it Chiang et al.}, 2001). The
data can be reproduced in models assuming that 50\% of the available
oxygen is in water ice. Edge-on disks offer a
special opportunity to study ices at mid-IR wavelengths in
absorption.  Examples are L1489 ({\it Boogert et al.}\ 2002), DG
Tau B ({\it Watson et al.}\ 2004) and CRBR2422.8-3423
({\it Thi et al.} 2002, {\it Pontoppidan et al.}
2005)(Fig.~3). Extreme care has to be taken in the interpretation of
these data since a large fraction of the ice features may arise in
foreground clouds. For the case of CRBR 2422.8-3423, a disk
viewed at an inclination of $\sim$70$^o$, comparison with nearby lines
of sight through the same core combined with detailed disk modeling
has been used to constrain the amount of ice in the disk.  H$_2$O ice
has an average line-of-sight abundance of $\sim 10^{-4}$ relative to
H$_2$, consistent with
significant freeze-out. CO$_2$ and CO ice are also
present, the latter only in the form where it is mixed with H$_2$O
ice.  The shape of the 6.85 $\mu$m ice band --usually ascribed to
NH$_4^+$ (e.g., {\it Schutte and Khanna}, 2004)-- shows evidence for heating
to 40--50 K, as expected in the warm intermediate layers of the
disk. Future studies of a large sample of edge-on disks can provide
significant insight into the abundance and distribution of ices in
disks, because the ice absorption depths, band shapes and feature
ratios depend strongly on the disk temperature structure and line of
sight, i.e., inclination ({\it Pontoppidan et al.}, 2005).

PAHs are important in the chemistry for at least three reasons: as
absorbers of UV radiation, as a heating agent for the gas, and as
potential sites of H$_2$ formation when classical grains have grown to
large sizes. Since they require UV radiation for excitation, PAHs are
also excellent diagnostics of the stellar radiation field and disk
shape (flaring or flat).  PAHs are detected in the spectra of at least
50\% of Herbig Ae stars with disks through 
emission features at 3.3, 6.2, 7.7, 8.6, 11.2, 12.8 $\mu$m ({\it Acke
and van den Ancker} 2004). For T Tauri stars, the features are weaker
and more difficult to see on top of the strong continuum, but at least
8\% of sources with spectral types later than F7 show the 11.2 $\mu$m
PAH feature ({\it Geers et al. 2006}, in prep.).
Ground-based
long-slit spectroscopy and narrow-band imaging at sub-arcsec resolution
has demonstrated that, at least for some disks, the PAH emission comes
from a region of radius 10--100 AU (e.g., {\it Geers et al.}, 2004,
{\it Habart et al.}, 2004).  The inferred PAH abundance is typically
$10^{-7}$ with respect to H$_2$, assuming that $\sim$10\% of the
carbon is in PAHs with 50-100 carbon atoms.  PAHs have been
detected in transitional disks, where
there is evidence for grain growth to $\mu$m size, albeit at a low
abundance of $10^{-9}$ ({\it Li and Lunine}, 2003, {\it Jonkheid et al.}, 2006).
This indicates that the PAHs and very small grains in the upper disk
layer may be decoupled dynamically from the larger silicate grains and
have a much longer lifetime toward grain growth and settling.

\subsubsection{Gas-phase molecules}

Vibration-rotation emission lines of gaseous CO at 4.7~$\mu$m are
detected toward a large fraction ($>$80\%) of T Tauri and Herbig Ae
stars with disks (e.g., {\it Najita et al.}, 2003, {\it Brittain and Rettig}, 2002,
{\it Blake and Boogert}, 2004). The CO lines can be excited by collisions in
the hot gas in the inner ($<$5--10 AU) disk as well as by IR or UV pumping
in the upper layers of the outer ($r > 10$ AU) disk. Searches for other molecular
emission lines have so far been largely unsuccessful except for the
detection of hot H$_2$O toward one young star, presumably arising in
the inner disk ({\it Carr et al.}, 2004).  A tentative detection of H$_3^+$ --
a potential tracer of protoplanets -- has been claimed toward one young
star (HD141569; {\it Brittain and Rettig}, 2002), but this remains unconfirmed 
({\it Goto et al.}, 2005).

\begin{figure*}
 \epsscale{1.7}
\plotone{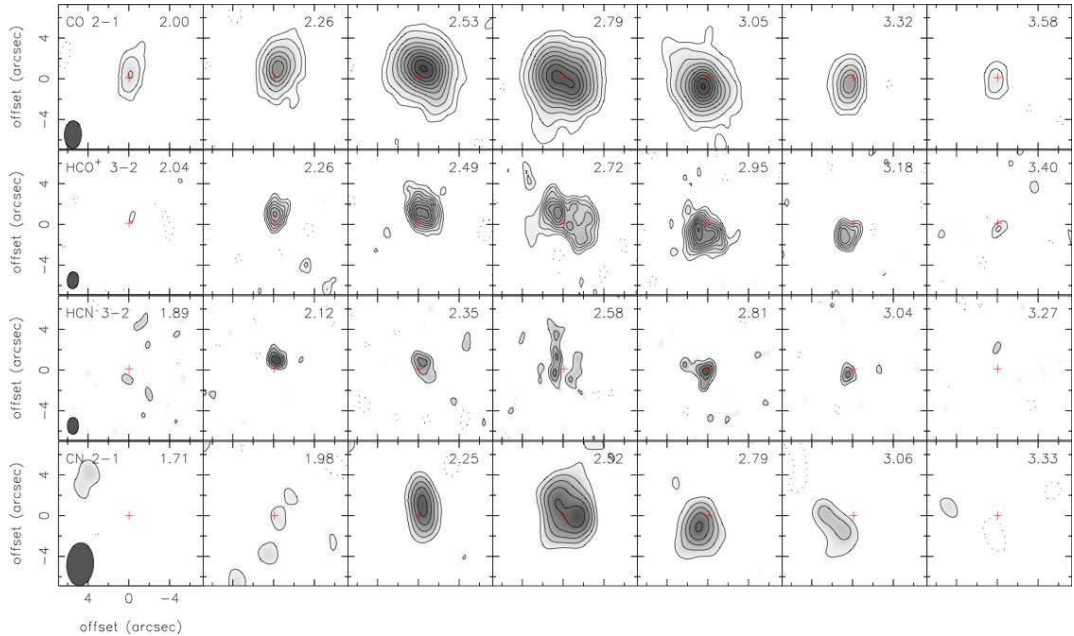}
\caption{SMA channel maps of the millimeter and submillimeter
spectral line emission from the TW Hya circumstellar disk. 
The ellipses at lower left in each series of panels
display the synthesized beam, which for the HCN 3-2 observations
achieves an effective spatial resolution of $\sim$60 AU.
Kindly provided by C. Qi in advance of publication ({\it Qi et al.}
2006, in prep.).}
\end{figure*}

Surprisingly, {\it Spitzer} has recently detected strong
absorption from C$_2$H$_2$, HCN, and CO$_2$ at $R=600$ toward one
young star with an edge-on disk, IRS 46 in Ophiuchus (Fig.~3)
({\it Lahuis et al.} 2006). The inferred abundances are high, $\sim
10^{-5}$ with respect to H$_2$, and the gas is hot, 300--700 K, with
linewidths of $\sim 20$ km s$^{-1}$. The most likely location for this
hot gas rich in organic molecules is in the inner ($<$few AU) disk,
with the line of sight passing through the puffed-up inner rim.  The
observed abundances are comparable to those found in inner disk
chemistry models (e.g., {\it Markwick et al.}\ 2002).

A novel method to probe the chemistry in the outer  ($r > 10$ AU) disk is through
observations of narrow, low-velocity [O I] $^1$D--$^3$P lines at 6300 \AA\
 showing signs of Keplerian rotation ({\it Bally et al.}, 1998; {\it Acke et
al.}, 2005).  The excited O $^1$D atoms most likely result from
photodissociation of OH in the upper flaring layers of the disk out to
at least 10 AU ({\it St\"orzer and Hollenbach}, 1998, {\it van Dishoeck and Dalgarno},
1984). The OH abundance required to explain the observed [O I] fluxes
is $\sim 10^{-7}-10^{-6}$, significantly larger than that found in
current chemical models.

The main reservoir of the gas in disks is H$_2$, which has its
fundamental rotational quadrapole lines at mid-IR
wavelengths. Searches for these lines have been performed by {\it Thi et
al.} (2001) with {\it ISO}, but tentative detections have not been
confirmed by subsequent ground-based data (e.g., {\it Richter et al.}, 2002,
{\it Sako et al.}, 2005), although H$_2$ vibrational lines have been detected
({\it Bary et al.}, 2003).  Constraining the amount of gas in disks is
important not only for Jovian planet formation, but also because the
gas/dust ratio affects the chemistry and thermal balance of the disk
as well as the dust dynamics.


\subsection{Millimeter- and Submillimeter-wave Spectroscopy}

At long wavelengths, where the dust emission is largely optically
thin, rotational line emission forms a powerful probe
of the physics and chemistry in disks. Indeed, at a
fiducial radius of 100 AU, with a temperature of 20-30 K, the
disk radiates preferentially at (sub)millimeter wavelengths.
Surveys of the mm-continuum and SEDs from disks ({\it Beckwith et al.},
1990) have confirmed the spatio-temporal properties of dust disks.
However, only a {\em handful} 
of objects have been investigated by detailed imaging, with
mm-wave interferometer observations of CO from disks around T Tauri
stars providing  among the earliest and most conclusive 
evidence for the expected $\sim$Keplerian velocity fields 
(see {\it Koerner et al.}, 1993, {\it Dutrey et al.}, 1994).

From such studies a suite of disks have been identified that
are many arc seconds in diameter, either because they are nearby
(TW Hya) or are intrinsically large (GM Aur, LkCa 15, DM Tau).
The age, large size, and masses of these disks
make them important for further study since they may represent 
an important transitional phase in which viscous disk spreading 
and dispersal competes with planetary formation processes. Their
large size makes them difficult, but feasible, targets for further 
chemical study. Work in this area began with the pioneering observations
of DM Tau and GG Tau with the IRAM 30m telescope ({\it Dutrey et al.}, 1997)
and TW Hya with the JCMT ({\it Kastner et al.}, 1997). Further detections
of the higher$-J$ lines of high dipole moment species such as HCN and
HCO$^+$ along with statistical equilibrium analyses demonstrated
that the line emission arises from the warm molecular layer with
$n_{\rm H_2} \approx$10$^6$--10$^8$ cm$^{-3}$,
$T \gtrsim$30 K ({\it van Zadelhoff et al.}, 2001). The very deepest
integrations have begun to reveal more complex species ({\it Thi et al.} 2004),
though the larger organics often seen toward protostellar hot cores
remain out of reach of existing single dish telescopes.

At present, aperture synthesis observations can only sense the
outer disk ($r>30-50$ AU, {\it Dutrey and Guilloteau}, 2004; see the Chapter by {\it Dutrey et al.})
for stars in the nearest
molecular clouds. Thus, the {\it chemical} imaging of disks is rarer still,
with studies concentrating on a few of the
best characterized T Tauri and Herbig Ae stars. Imaging studies of
LkCa 15, for example, have detected a number of isotopologues of
CO along with the molecular ions HCO$^+$ and N$_2$H$^+$ and the more
complex organics formaldehyde and methanol ({\it Duvert et al.}, 2000,
{\it Aikawa et al.}, 2003, {\it Qi et al.}, 2003).  For this disk at least,
molecular depletion of molecules onto the icy mantles of 
grains near the disk midplane is found to be extensive, but the
fractional abundances and ionization in the warm molecular layer are in line with
those seen toward dense PDRs (\S 2.1).  While the lines from the less 
abundant species can be detected, they were too weak to image
with good signal-to-noise. Thus, while millimeter-wave rotational
line emission is a good tracer of the outer disk velocity field it
is not a robust tracer of the mass unless the chemistry is very
well understood.

New observational facilities are poised to change this situation
dramatically, as illustrated by the recent SMA results on the TW Hya
disk presented in Figure 4 ({\it Qi et al.} 2006, in prep.). At a
distance of only 56 pc, observations of this source provide nearly
2--3 times the effective linear resolution of studies in Taurus and
Ophiuchus. Thus, channel maps such as those presented can be used to
derive a great deal about the physical and chemical structure of the
disk - its size and inclination ({\it Qi et al.} 2004), the run of
mass surface density and temperature with radius, the chemical
abundance ratios with radius in the outer disk, etc. Ongoing
improvements to existing arrays such as the eSMA, PdBI, and CARMA will
enable similar studies for a large number of disks in the near future,
and will push the radii over which chemical studies can be pursued
down to 10-20 AU.  Resolving the chemical gradients discussed in \S4.3
and 4.4 and studying the chemistry in the 1-10 AU zone of active
planet formation will require even greater sensitivity and spatial
resolution, and awaits ALMA.

\section{\textbf{CHEMICAL AND PHYSICAL LINKS}}

Protoplanetary disks are evolving in many different, but connected, ways.
Small micron sized dust grains collide, coagulate, collisionally
fragment, and settle
in a process that ultimately produces planets ({\it Weidenschilling}, 1997).
At the same time the disk is
viscously evolving, with indications that the disk mass accretion rate decreases
with age ({\it Hartmann et al.}, 1998).  Finally, disks evolve chemically, with the
eventual result that all heavy elements are frozen on grains in the midplane and, prior to
gas dissipation, 
chemistry consists of 
H$_2$ ionization and the deuteration sequence to H$_2$D$^+$,
D$_2$H$^+$, and D$_3^+$ (see \S 4.3 and \S 5).
The interconnections between these types of evolution is, at
present, poorly understood. Nonetheless, some physical and chemical connections
have become clear, which we review here.

\subsection{Grain Evolution}
\label{sec:grainevol}
\begin{figure*}
\centering
\includegraphics[width=12cm]{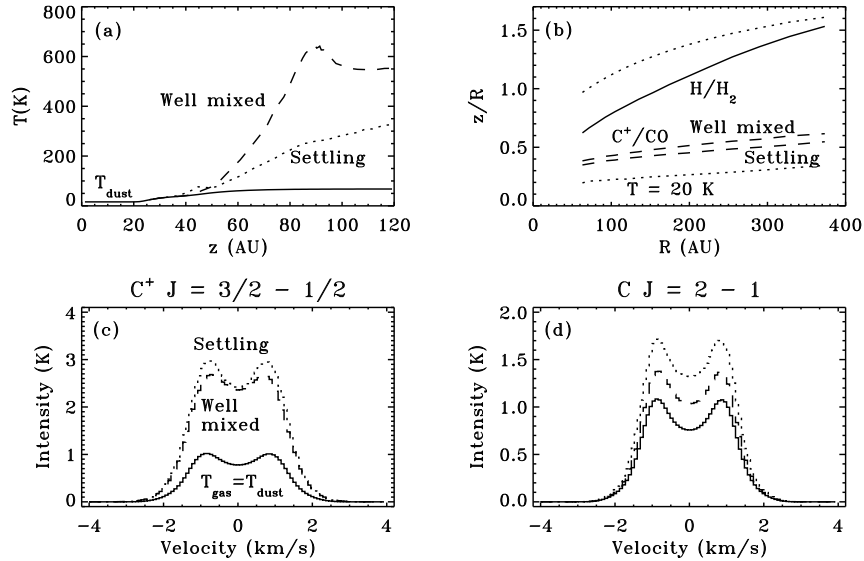}
\caption{
(a) Vertical distribution of gas (well mixed model: dashed line; settling model:
dotted line) and dust temperatures (solid line).  (b) Vertical chemical structure
with the edge of the disk and 20 K isotherm as dotted lines, the H/H$_2$ transition
as a solid line and the C$^+$/CO transitions in dashed lines (for both well mixed and
settled models). (c) C$^+$ 158 $\mu$m emission and (d) C 370 $\mu$m emission lines
for $T_g = T_d$ (solid), and $T_g$ calculated independently for well mixed (dashed)
and settling (dotted) models.
Taken from {\it Jonkheid et al.} (2004). 
}
\label{fig:grainevol}
\end{figure*}

The onset of grain evolution within a protoplanetary disk
consists of collisional growth of sub-micron sized
particles into larger grains; the process continues until the larger grains
decouple from the gas and settle to an increasingly dust-rich midplane
({\it Nakagawa et al.}, 1981; {\it Weidenschilling and Cuzzi},
1993; {\it Beckwith et al.}, 2000).
Evolving dust grains within irradiated
disks reprocess stellar and accretion-generated UV and optical photons into
the IR and sub-mm.
Thus, the dust thermal emission spectrum bears information on the
grain size distribution and spatial location.  This, together with the
spectral features discussed in \S 3.1.1, is used to provide
evidence for grain evolution in $\sim$1 Myr T Tauri systems
({\it Beckwith et al.} 2000, see the Chapter by {\it  Dullemond et al.}, and references therein).

Grain coagulation can alter the chemistry through the reduction in the total
geometrical cross-section, lowering the adsorption rate and the Coulomb
force for ion-electron grain recombination.
Micron-sized grains couple to the
smallest scales of turbulence ({\it Weidenschilling and Cuzzi}, 1993) and have a thermal,
Brownian, velocity distribution. Thus, the timescale of grain-grain collisions is,
$\tau_{gr-gr} \propto a_d^{5/2}/(T_d^{1/2}\xi n_{\rm H})$, where $a_d$ is the grain radius, $\xi$
the gas-to-dust mass ratio, and $T_d$ the dust temperature ({\it Aikawa et al.}, 1999).
In this fashion grain coagulation proceeds faster at small radii where the
temperatures and densities are higher.
{\it Aikawa et al.}, (1999) note that the longer
timescale for adsorption on larger grains leaves more time for
gas-phase reactions to drive toward a
steady-state solution; this involves more carbon trapped in CO 
as opposed to other more complex species.


Overall, the evolution of grains, both coagulation and sedimentation,
can be a controlling factor for the chemistry.  As grains grow the
UV opacity, which is dominated by small grains,
decreases, allowing greater penetration of ionizing/dissociating photons (e.g. {\it Dullemond and Dominik} 2005).
As an example, in the coagulation models of {\it Dullemond and Dominik}, (2005) the
integrated vertical UV optical depth at 1 AU decreases over several orders of
magnitude, towards optically thin over the entire column
(see also {\it Weidenschilling}, 1997).
The chemical effects are demonstrated in
Fig.~\ref{fig:grainevol}(b) where we show the C$^+$ to CO transition as a
function of vertical distance for two dust models, well-mixed and settled
({\it Jonkheid et al.} 2004).   In the settled model the CO
transition occurs at slightly smaller height, even though this model assumes
PAH's remain present in the upper atmosphere.
Thus as grains evolve there will be a gradual
shifting of the warm molecular layer deeper into the disk, eventually into the
midplane.  Because the grain emissivity, density, and temperature will also change, the
chemical and emission characteristics of this layer may be altered ({\it Aikawa and
Nomura}, 2006).
These effects are magnified in the inner disk, where there is
evidence for significant grain evolution in a few systems ({\it Uchida et al.}, 2004;
{\it Calvet et al.}, 2005) and deeper penetration of energetic radiation ({\it Bergin
et al.}, 2004).
A key question in this regard is the number of small grains (e.g. PAHs)
present in the atmosphere of the
disk during times when significant coagulation and settling has occurred
(e.g. {\it Dullemond and Dominik}, 2005; {\it D'Alessio et al.}, 2006).

It is worth noting that the penetration of X-ray photons is somewhat different
than those at longer UV wavelengths.  Absorption of UV radiation is dominated
by the small grains, while X-rays are absorbed at the atomic scale by heavy
metals predominantly trapped in the grain cores.  Thus, coagulation will have a
greater effect on UV photons.  When the grain mass is
distributed downwards by settling, the X-ray penetration depth increases,
eventually to the limit of
total heavy element depletion where the opacity at 1 keV decreases by a factor
of $\sim 4.5$ (i.e. the H and He absorption limit; {\it Morrison and McCammon}, 1983).



\subsection{Gas Thermal Structure}

Models of disk physical structure have generally assumed that dust and
gas temperatures are in equilibrium ({\it Chiang and Goldreich}, 1997; {\it D'Alessio et
al.}, 1998).
However, the disk vertical structure is set by the temperature of the dominant
mass component, hydrogen, which under some conditions in the upper disk
atmosphere is thermally decoupled from dust ({\it Chiang and
Goldreich}, 1997).  

Irradiated disk surfaces are analogs to
interstellar PDRs, which have a history of detailed thermal balance calculations
(see {\it Hollenbach and Tielens}, 1999 and references therein).
In the studies of the gas thermal balance in disk atmospheres ({\it Jonkheid et al.} 2004; {\it Kamp and Dullemond} 2004;
{\it Nomura and Millar} 2005) a number of heating mechanisms have been investigated:
including photoelectric heating by PAHs and large grains,
UV excitation of H$_2$ followed by collisional de-excitation, H$_2$ dissociation,
H$_2$ formation, gas-grain collisions, carbon ionization, and cosmic rays.
In decoupled layers,
the gas cools primarily by atomic ([O I], [C II], [C I]) and molecular
(CO) emission, with the dominant mechanism a function of radial and vertical distance.

\begin{table*}
\begin{tabular*}{6.8in}[t]{lllll}
\multicolumn{5}{c}{TABLE 1} \\
\multicolumn{5}{c}{DISK IONIZATION PROCESSES AND VERTICAL ION STRUCTURE} \\
\hline
\hline
\multicolumn{1}{c}{Layer/Carrier} &
\multicolumn{1}{c}{Ionization Mechanism} &
\multicolumn{1}{c}{$\Sigma_{\tau = 1}$ (g cm$^{-2}$)\tablenotemark{a}} &
\multicolumn{1}{c}{$\alpha_{r}$ (cm$^{-3}$s$^{-1}$)\tablenotemark{b}} &
\multicolumn{1}{c}{$x_e^{c}$} \\\hline
{\bf Upper Surface} & UV photoionization of H$^{d}$ &  6.9 $\times 10
^{-4}$
& $\alpha_{\rm{H}^+} = 2.5 \times 10^{-10}T^{-0.75}$ & $> 10^{-4}$\\
H$^+$ & $k_{\rm{H}^+} \sim 10^{-8}$ s$^{-1}$ &&\\\hline
{\bf Lower Surface} & UV photoionization of C$^{e}$ & 1.3 $\times 10^
{-3}$ &
$\alpha_{\rm{C}^+} =1.3 \times 10^{-10}T^{-0.61}$ & $\sim 10^{-4}$ \\
C$^{+}$ & $k_{\rm{C}^+} \sim 4 \times 10^{-8}$ s$^{-1}$ &&\\\hline
{\bf Warm Mol.} & Cosmic-$^{f}$ and X-Ray$^{g}$ Ionization
& 96 (CR) & $\alpha_{\rm{H_3}^+}=-1.3 \times 10^{-8} +$& $10^{-11\rightarrow -6}$  \\
H$_3^+$,HCO$^+$ & $\zeta_{cr} = \frac{\zeta_{cr,0}}{2}[{\rm exp}(-\frac{\Sigma_1}{\Sigma_{cr}}) +
{\rm exp}(-\frac{\Sigma_2}{\Sigma_{cr}})]$     & & 1.27 $\times 10^{-6}T^{-0.48}$ &
 \\


& $\zeta_X = \zeta_{X,0}
\frac{\sigma(kT_X)}{\sigma(1keV)}L_{29}J(r/{\rm AU})^{-2}$ & 0.008 (1keV) &  $\alpha_{\rm{HCO}^+} = 3.3 \times
10^{-5}T^{-1}$ &  \\

& & 1.6 (10 keV) &  &  \\\hline

{\bf Mid-Plane} & Cosmic-ray$^{f}$ and Radionuclide$^{h}$&
 & & \\
& $\zeta_{R} = 6.1 \times 10^{-18}$ s$^{-1}$  & & &
 \\
Metal$^{+}$/gr & ($r < 3$ AU) & &
$\alpha_{\rm{Na}^+} = 1.4 \times 10^{-10}T^{-0.69}$ & $<
10^{-12}$ \\
HCO$^{+}$/gr & ($3 < r < 60$ AU) & & $\alpha_{gr}$ (see text) & 10$^{-13,-12}$ \\
H$_3^{+} -$ D$_3^+$ & ($r > 60$ AU) & & $\alpha_{\rm{D_3}^+} = 2.7 \times
10^{-8}T^{-0.5}$ & $> 10^{-11}$\\\hline
\vspace{-7mm}
\tablenotetext{a}{Effective penetration depth of radiation (e.g. $\tau = 1$ surface).}
\tablenotetext{b}{
Recombination rates from UMIST database ({\it Le Teuff et al.},
2000), except for
H$_3^+$ which is from {\it McCall et al.}, (2004) and D$_3^+$ from {\it Larsson et al.}, (1997).}
\tablenotetext{c}{
Ion fractions estimated from {\it Semenov et al.} (2004) and {\it Sano et
al.}, (2000). Unless noted values are relevant for all radii.}
\tablenotetext{d}{Estimated at 100 AU assuming 10$^{41}$ s$^{-1}$
ionizing photons ({\it Hollenbach et al.}, 2000) and  $\sigma = 6.3 \times
10^{-18}$ cm$^{2}$ (H photoionization cross-section at threshold).
This is an overestimate as we assume all ionizing photons are at the
Lyman limit.
}
\tablenotetext{e}{Rate at the disk surface at 100 AU using the
radiation field from {\it Bergin et al.} (2003).
}
\tablenotetext{f}{Taken from {\it Semenov et al.} (2004).  $\zeta_{cr,0} = 1.0 \times 10^{-17}$ s$^{-1}$ and
$\Sigma_1(r,z)$ is the surface density above the point with height $z$ and
radius
$r$ with $\Sigma_2(r,z)$ the surface density below the same point.  $\Sigma_{cr} =
96$ g
cm$^{-2}$ as given above ({\it Umebayashi and Nakano}, 1981).
}
\tablenotetext{g}{X-ray ionization formalism from {\it Glassgold et al.} (2000).  $\zeta_{X,0} = 1.4 \times10^{-10}\;s^{-1}$, while
$L_{29} = L_X/10^{29}$ erg s$^{-1}$ is the X-ray luminosity and $J$ is an attenuation factor, $J =
A\tau^{-a}e^{-B\tau^b}$, where $A$ = 0.800, $a$ = 0.570, $B$ = 1.821, and $b$ = 0.287 (for energies around 1
keV and solar abundances).
}
\tablenotetext{h}{$^{26}$Al decay from {\it Umebayashi and Nakano} (1981). If $^{26}$Al is not present $^{40}$K dominates with $\zeta_R = 6.9 \times 10^{-23}$ s$^{-1}$.}
\end{tabular*}
\label{tab:ion}
\end{table*}

A sample of these results are shown in
Fig.~\ref{fig:grainevol}; note that the gas temperature
can exceed that of the dust in the upper atmosphere
(Fig.~\ref{fig:grainevol}a), which 
has consequences for the gas phase emission (Fig.~\ref{fig:grainevol}c,d).
The inclusion of PAHs into the models has an
effect as PAHs provide additional heating power and are strong UV
absorbers.  Thus, grain evolution can significantly alter the thermal structure (see
Fig.~\ref{fig:grainevol}) by removing PAHs and small grains through coagulation and
larger grains by settling, reducing photoelectric
heating.
In disks where grains have grown to micron size, and which are optically
thin to UV radiation, other heating processes such as the drift velocity
between the dust and gas may become important
(see {\it Kamp and van Zadelhoff}, 2001).


One of the largest chemical influences for most of the disk mass
is the freeze-out of molecular species
onto grain surfaces.  In general, the loss of gas coolants would produce a temperature
rise, but in (mid-plane) layers dominated by freeze-out, the densities are high enough
to thermally couple the gas to the dust (see, e.g. {\it Goldsmith}, 2001).

\subsection{The Ionization Fraction}
\label{sec:ionfrac}

Over the past few years the disk fraction ionization has received a high degree of
attention owing to the appreciation of the Magneto-Rotational Instability (MRI) as
a potential mechanism for disk angular momentum transport (e.g. {\it Balbus and
Hawley}, 1991; {\it Stone et al.}, 2000).
The chemical evolution is linked to dynamics as the presence of ions
is necessary to couple the gas to the magnetic field.
As we will outline, most of the ionization processes are active at the surface
and there exists the
potential that accretion may only be active on the surface ({\it Gammie}, 1996, but
see also {\it Klahr and Bodenhiemer}, 2003; {\it Inutsuka and Sano}, 2005).

\begin{figure*}
\includegraphics[width=15cm]{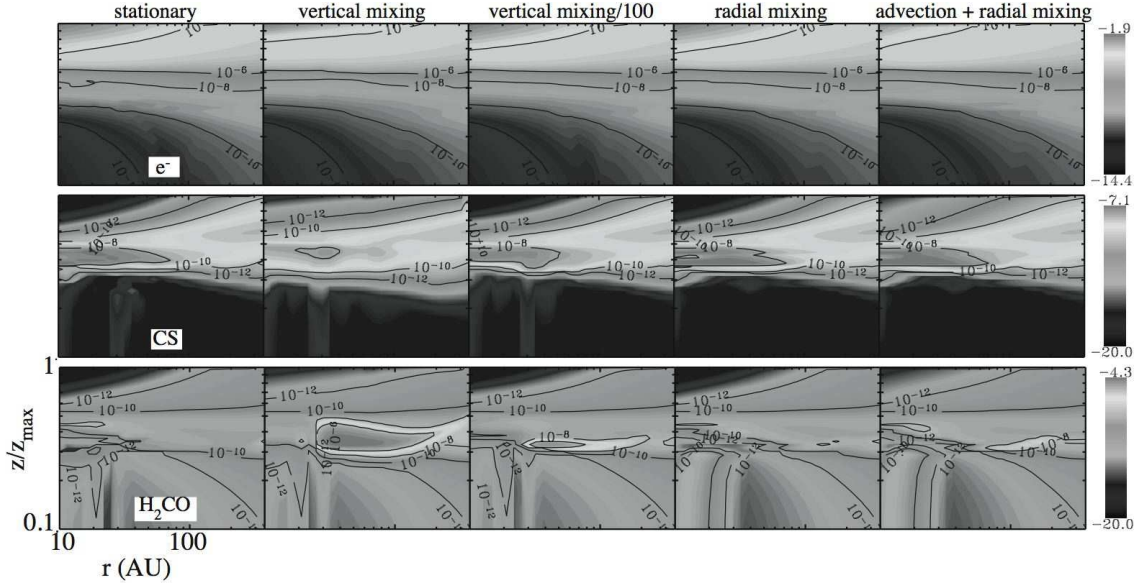}
\caption{ Shown are molecular abundances in the 2D flared disk of {\it
D'Alessio et al.} (1999) after $10^6$ years of the evolution.  The
x-axis represents radii from 10 to 370 AU on logarithmic scale, while
y-axis corresponds to normalized vertical extend of the disk,
$z/z_{max}$, that ranges from 0.1 to 1.0 on logarithmic scale. The
abundance values are given in respect to the total amount of hydrogen
nuclei.  Several models with variable mixing were considered and are
labeled above each panel (note: vertical mixing/100 is a model with
reduced mixing).  Kindly provided by D. Semenov in advance of
publication ({\it Semenov et al.}, 2006, in prep.).  }
\label{semenov}
\end{figure*}

In equilibrium, the ion fraction, $x_e = n_e/n_{\rm H}$, can be expressed by:
$x_e = \sqrt{\zeta/(\alpha_{r} n_H})$,
where $\zeta$ is the ionization rate and $\alpha_{r}$ the
electron recombination rate.
In Table~1 we provide an overview of disk ionization.
The top left panel in Fig.~\ref{semenov} also shows the electron
abundance (equivalent to the ionization fraction) from a detailed
model ({\it Semenov et al.}, 2006, in prep.).  This figure, along with
Table~1, can be used as a guide for the following discussion.  For a
discussion of the validity of the equilibrium assumption see {\it
Semenov et al.} (2004), while {\it Ilgner and Nelson} (2005a) provide
a detailed comparison of ionization and MRI for a variety of reaction
networks.

The observed FUV radiation excess
produces high ion fractions,
but only
over a small surface column, with C$^+$ as the charge carrier
(the ionized hydrogen layer will be quite small).
Deeper inside the warm molecular layer is reached, where
primary charge carriers are molecular ions
produced by X-ray ionization of H$_2$ ({\it Glassgold et al.}, 1997)
and, when this decays, cosmic ray
ionization.
It is worth noting that X-ray flares are observed in T Tauri systems
(e.g. {\it Favata et al.}, 2005), after which
there will exist a burst of ionization on the disk surface that will
last for $\tau_{r} \sim 1/(\alpha_{r} n_e)$.

An important question is whether cosmic rays penetrate the inner disk.
Within our own planetary system, the Solar wind excludes ionizing
cosmic rays.  Estimates of mass loss rates from young star winds
significantly exceed the Solar mass loss rate ({\it Dupree et al.},
2005), and may similarly exclude high energy nuclei.
In the case of cosmic ray penetration, primary charge carriers range
from metal ions and/or grains at
small radii ($\sim$ 1 AU) to molecular ions for $r >$ few AU with $x_e \sim
10^{-13}$ near the midplane.
If ionizing cosmic rays are excluded, radionuclides can produce $x_e \sim 10^{-3}(T/20~{\rm K})^{-0.5}/\sqrt{n_{\rm H}}$ (assuming H$_3^+$ as the dominant ion and $^{26}$Al is present; if $^{26}$Al is not present the pre-factor is 10$^{-8}$).
The presence or absence of metal ions in the midplane can also affect MHD driven dynamics
({\it Ilgner and Nelson}, 2005b).   Metal ions have recombination
times longer than molecular ions or the diffusive timescale and, if present, can be
important charge reservoirs.

In dense protostellar cores models and observations now suggest near total
freeze-out of heavy species that results in D$_3^+$ and other forms of
deuterated H$_3^+$ becoming
important charge carriers ({\it Roberts, Herbst, and Millar}, 2004; {\it Walmsley, Flower,
and Pineau des For\^ets}, 2004).  The disk midplane should present a similar
environment. The recent detection of H$_2$D$^+$ by {\it Ceccarelli
et al.} (2004) in the outer disks of TW Hya and DM Tau, supports this view.

In the midplane one issue is
the electron sticking coefficient ($S_e$) to grains.
{\it Sano et al.} (2000)
assume $S_e =$ 0.6 based on the work of {\it Nishi et al.}
(1991), while {\it Semenov et al.} (2004) assume a
strong temperature dependence with $S_e$ essentially zero at high temperatures.  Thus
in the inner disk the primary charge carrier differs between these two models with
grains dominating in the {\it
Sano et al.} (2000) model and
molecular/metal ions in the {\it Semenov et al.} (2004) model (for a discussion of
electron sticking coefficients see {\it Weingartner and Draine}, 2001).
If grains are the dominant charge carrier the recombination rate $\alpha_{gr}$ is the grain
collisional timescale with a correction for long-distance Coulomb focusing:
$\alpha_{gr} = \pi a_d^2 n_{gr}v(1 + e^2/ka_dT_d)$. 
At $T_d = 20$ K {\it Draine and Sutin} (1987)
show that for molecular ions, grain recombination will dominate when
$n_e/n_H < 10^{-7}(a_{min}/3$\rm{\AA}$)^{-3/2}$.
Grains can be positive or negative and carry multiple charge: {\it Sano et al.} (2000) find that the
total grain charge is typically negative, while the amount of charge is 1-2e$^-$, varying with
radial and vertical distance.
Since the criteria for the MRI instability include the mass of the charged
particles, it is important to determine whether grains or molecules
are the primary charge carriers.

\subsection{Mixing}

Chemical mixing within the solar nebula has a long history due to
important questions regarding the potential transport of material from warm, thermochemically
active, regions (either inner solar nebula or Jovian sub-nebula) into colder
inactive regions ({\it Lewis and Prinn}, 1980, {\it Prinn and Fegley}, 1981).
It is clear that some movement of processed material is likely, for example
the detection of crystalline silicates in comets ({\it Crovisier et al.}, 1997; {\it
Bockel{\'e}e-Morvan et al.}, 2000)
and the chondritic refractory inclusions in meteorites ({\it MacPherson et al.},
1988) implies some mixing.
However, the mixing efficiency has been a matter of debate
({\it Stevenson}, 1990; {\it Prinn}, 1990; 1993).

In terms of the dynamical movement of gas within a protoplanetary disk and its
chemical effects, a key question is whether the chemical timescale, $\tau_{chem}$, is less
than the relevant dynamical timescale, $\tau_{dyn}$, in which case  the chemistry will
be in equilibrium and unaffected by the motion.
If $\tau_{dyn} <  \tau_{chem}$ then mixing will alter
the anticipated composition.  These two constraints are the equilibrium and
disequilibrium regions (respectively) outlined in {\it Prinn} (1993).
What is somewhat different in our current perspective is the recognition of an active
gas-phase chemistry on a photon-dominated surface.
This provides another potential mixing reservoir in the vertical direction, as
opposed to radial, which was the previous focus.


It is common to parameterize the transfer of angular momentum
in terms of the turbulent viscosity,
$\nu = \alpha c_s H$,
where $\nu$ is the viscosity, $c_s$ the sound speed, $H$ the disk scale
height, and $\alpha$ is a dimensionless parameter
({\it Shakura and Sunyaev}, 1973; {\it Pringle}, 1981).
{\it Hartmann et al.} (1998) empirically constrained the $\alpha$-parameter to be
$\lesssim
10^{-2}$ for a sample of T Tauri disks.

 The radial disk viscous timescale is
$\tau_{\nu} = r^2/\nu$ or,

\[
\tau_{\nu} \sim 10^{4}{\rm yr}\left(\frac{\alpha}{10^{-2}}\right)^{-1}
\left(\frac{T}{100\,{\rm K}}\right)^{-1}\left(\frac{r}{1\,{\rm AU}}\right)^{\frac{1}{2}}
\left(\frac{M_*}{M_{\odot}}\right)^{\frac{1}{2}}.
\]

\noindent The diffusivity, $D$, is not necessarily the same as the
viscosity, $\nu$ (e.g. {\it Stevenson}, 1990), even though some
treatments equate the two ({\it Ilgner et al.}, 2004; {\it Willacy et
al.}, 2006).  In the case of MRI, {\it Carballido et al.} (2005)
estimate $\nu/D \sim 11$, i.e., turbulent mixing is much less efficient
than angular momentum transport (but, see also {\it Turner et al.}
2006 where $\nu/D \sim 1-2$).  With knowledge of the $\nu/D$ ratio,
the above equation for $\tau_{\nu}$ serves as an estimate of the
dynamical timescale for diffusion.


Recent models that include dynamics with chemistry generally can be
grouped into two categories. Ones that include only advection ({\it
Duschl et al.}, 1996; {\it Finocchi et al.}, 1997; {\it Willacy et al.},
1998; {\it Aikawa et al.}, 1999; {\it Markwick et al.}, 2002) and ones
that investigate the effects of vertical and/or radial mixing
including advection ({\it Wehrstedt and Gail}, 2003; {\it Ilgner et
al.}, 2004; {\it Willacy et al.}, 2006; {\it Semenov et al.}, 2006, in
prep.).  The effects of advection on the chemical evolution are
dominated by migration of icy grains towards the warmer inner disk
where the volatile ices evaporate.  Most of the species that desorb
are processed via the active gas phase and/or gas-grain chemistry into
less volatile species that freeze out (e.g. {\it Aikawa et al.}, 1999).
For instance, N$_2$ evaporates at $T> 20$~K (\"{O}berg et al. 2005)
and is converted to HCN, and other less volatile species, via
ion-molecule reactions.
This would predict a strong temporal dependence on the chemistry in the accreting material 
within the evaporative zones.

\setcounter{table}{1}
\begin{deluxetable}{l l l c l}
\tabletypesize{\small}
\tablecaption{D/H ratios in comets, disks and cores \label{D/H}}
\tablewidth{0pt}
\tablehead{
\colhead{Region Type} &
\colhead{Object} &
\colhead{Species} &
\colhead{D/H ratio} &
\colhead{Reference}}
\startdata
Hot cores & various & HDO & $3.0\times 10^{-4}$ & {\it Gensheimer et al.}, (1996) \\
          & various & DCN    & $0.9-4.0\times 10^{-3}$ & {\it Hatchell et al.}, (1998)\\
Low-mass protostars & IRAS 16293-2422 & HDO & $3.0 \times 10^{-2}$ & {\it Parise et al.}, (2005) \\
               &    IRAS 16293-2422 & CH$_3$OD & $2.0 \times 10^{-2}$ & {\it Parise et al.}, (2002) \\
               &   IRAS 16293-2422 & CH$_2$DOH & $3.0  \times 10^{-1}$ & {\it Parise et al.}, (2002) \\
               & various & HDCO & $5.0-7.0  \times 10^{-2}$ & {\it Roberts et al.}, (2002) \\
               & various & NH$_2$D & $1.0-2.8  \times 10^{-1}$ & {\it Roueff et al.}, (2005) \\
               & various & DCO$^+$ & $1.0 \times 10^{-2}$ & Class 0 average; {\it J{\o}rgensen et al.}, (2004)\\
               & various & DCN & $1.0  \times 10^{-2}$  & Class 0 average; {\it J{\o}rgensen et al.}, (2004)\\
Dark Cores & L1544 & DCO$^+$ & $4.0  \times 10^{-2}$ & {\it Caselli et al.}, (2002)\\
           & L134N & DCO$^+$ & $1.8  \times 10^{-1}$ & {\it Tin\'e et al.}, (2000)\\
           & TMC-1 & DCN & $2.3  \times 10^{-2}$ &  {\it van Dishoeck et al.}, (1993)\\
Disks  & TW Hya & DCO$^+$ & $3.5 \times 10^{-2}$ & {\it van Dishoeck et al.}, (2003) \\
       & LkCa 15 & HDO     & $6.4  \times 10^{-2}$ & {\it Kessler et al.}, (2003) \\
       & DM Tau & HDO &      $1.0  \times 10^{-3}$ & {\it Ceccarelli et al.}, (2005), {\it Dominik et al.}, (2005) \\
       & LkCa 15 & DCN & $< 0.002$ & {\it Kessler et al.}, (2003) \\
Comets & Halley & HDO & $(3.2\pm 0.3)\times 10^{-4}$ & {\it Eberhardt et al.}, (1995)\\
       & Hyakutake    & HDO &$(2.9\pm 1.0) \times 10^{-4}$ & {\it Bockel{\'e}e-Morvan et al.}, (1998) \\
       & Hale-Bopp & HDO & $(3.3\pm 0.8) \times 10^{-4}$ & {\it Meier et al.}, (1998)\\
       & Hale-Bopp & HDO & $2 \times 10^{-3}$ & {\it Blake et al.}, (1999) \\
       & Hale-Bopp & DCN & $(2.3\pm 0.4) \times 10^{-3}$ & {\it Meier et al.} (1998) \\
\enddata
\vspace{-8mm}
\tablecomments{\footnotesize
HDO has only been {\em tentatively} detected in disks, whereas H$_2$O has not. 
In these cases
({\it Ceccarelli et al.}, 2005; {\it Kessler et al.}, 2003) 
the D/H ratio is estimated by model calculation
and is therefore highly uncertain.
}
\end{deluxetable}

As an example of more complex dynamical models, Fig.~\ref{semenov} shows
results from {\it Semenov et al.} (2006, in preparation), but see also {\it Willacy et al.}
(2006). For these models, radial and vertical
mixing is incorporated
into a disk model that also includes all other relevant physical/chemical
processes described in \S 1.2.  The model assumes a traditional alpha-disk
that is fully dynamically active with $\nu/D = 1$.
This model reproduces the basic structure shown in
Fig.~\ref{overview}, with the presence of a warm molecular layer and a drop in abundance
towards the mid-plane (most dramatically seen in CS).  When vertical transport and/or
radial transport is included the essential features of the basic structure are
preserved, in the sense that the warm molecular layer still exists, although it
may be expanded ({\it Willacy et al.} 2006).  This is readily understood as the
chemical timescales driven by the photodissociation, $\tau_{\rm chem} \lesssim
100-1000 (r/100\;\rm{AU})^{2}$ yrs,
are less than the dynamical timescale at all radii, thus the photo-chemical
equilibrium is preserved.
The effects of radial mixing and advection are mostly important for the upper disk
atmosphere.  Two results stand out.
(1) The electron abundance structure shows little overall
change.  Thus ionization equilibrium is preserved throughout the disk.
(2) There is clear evidence for abundance enhancements of key species (such as
H$_2$CO) in models with vertical mixing.  The molecules with the largest
abundance enhancements in
the vertical mixing models are those that are more volatile than water, but are also
major components of the grain mantle.  This suggests
that observations of these species may ultimately be capable of constraining disk mixing.

However, all predictions are highly uncertain.  If MRI powers accretion, and cosmic rays
do not penetrate to the midplane, then the bulk of the disk mass will not participate in
any global mixing.  
Moreover, the outer disk may be actively turbulent while the inner
disk may be predominantly quiescent (excluding the surface ionized by X-rays).   Thus,
while current models are suggestive of the importance of mixing for the disk
chemical evolution, significant questions remain.

\section{\textbf{Deuterated Species in Disks and Comets}}

Isotopic fractionations are measured in primordial materials in comets
and meteorites, and considered to be good tracers of their origin and
evolution.  Here we review recent progress on deuterium fractionation
in relation to comets (see the Chapter by Yurimoto et al.\ for a
discussion of oxygen fractionation).  In Table~2 we list molecular D/H
ratios, defined as $n$(XD)/$n$(XH), observed in protoplanetary disks,
comets and low-mass cores.


The two-dimensional $(r-z$) distributions of deuterated species at $r\ge 26$ AU
have been calculated in models of {\it Aikawa and Herbst} (2001) and {\it Aikawa et al.}
(2002). At low temperatures D/H fractionation proceeds via
ion-molecule reactions;
species such as H$_3^+$ and CH$_3^+$ are enriched in deuterium because of
the difference in zero-point energies between isotopomers and rapid exchange
reactions such as H$_3^+$ + HD $\to$ H$_2$D$^+$ + H$_2$ (e.g. {\it Millar et al.},
1989). Since CO is the dominant reactant with H$_2$D$^+$, CO
depletion further enhances the H$_2$D$^+$/H$_3^+$ ratio. The deuterium
enrichment propagates to other species via chemical reactions (see the Chapters by 
{\it Ceccarelli et al.} and {\it Di Francesco et al.}).
Hence the D/H ratios of HCN
and HCO$^+$ tend to increase towards the midplane with low temperature
and heavy molecular depletion, while their absolute abundances reach the
maximum value at some intermediate height. The column density D/H ratios of
HCO$^+$, HCN and H$_2$O integrated in the vertical direction are $10^{-2}$
at $r\gtrsim 100$ AU and $10^{-3}$ at 26 AU $ \lesssim r \lesssim 100$ AU
in {\it Aikawa et al.} (2002). H$_3^+$ and its deuterated families
(H$_2$D$^+$, HD$_2$H$^+$ and D$_3^+$), on the other hand, are abundant
in the midplane, and the D/H ratio can even be higher than unity.  At
present, the data on D/H ratios in disks (with both deuterated and
hydrogenated species observed) is limited to the detection of DCO$^+$
({\it van Dishoeck et al.}, 2003), which is in agreement with models
({\it Aikawa et al.}, 2002).


In contrast to the mm observations of gas in the outer disk, comets
carry information on ice (rather than gas) at radii of $5-30$ AU
(e.g. {\it Mumma et al.}, 1993).  The similarity in molecular D/H
ratios between comets and high-mass hot cores has been used to argue
for an interstellar origin of cometary matter, but the D/H ratios in
low-mass star-forming regions (e.g. TMC-1) are higher than those in
comets, casting questions as to the interstellar origin scenario
(e.g. {\it Irvine et al.}, 1999). In recent years hot cores are found
around low- or intermediate-mass protostars (see IRAS 16293-2422).  In
addition, temporal and spatial variation in molecular D/H ratios are
found in low-mass dense cores, where the bulk of ices formed (e.g.
{\it Bacmann et al.}, 2003; {\it Caselli}, 2002). Hence, what one
means by ``interstellar'' is ambiguous. Eventually, molecular
evolution from cores to disks and within disks should be investigated.

{\it Aikawa and Herbst} (1999) calculated molecular abundances and D/H
ratios in a fluid parcel accreting from a core to the disk, and then
from the outer disk radius to the comet-forming region (30 AU),
showing that ratios such as DCN/HCN depend on the ionization rate in
the disk, and can decrease from 0.01 to 0.002 (if the migration takes
$10^6$ yr and the ionization rate is 10$^{-18}$ s$^{-1}$) due to
chemical reactions during migration within the disk. This model,
however, assumed that the fluid parcel migrates only inward within the
cold ($T\lesssim 25$ K) midplane, which results in the survival of
highly deuterated water accreted from the core. {\it Hersant et al.}
(2001) solved the diffusion equation to obtain the D/H ratio in the
disk; initially high D/H ratios of H$_2$O and HCN are lowered by
mixing with the poorly deuterated material from the smaller
radii. This model, however, considered only thermal reactions. In
reality deuterium fractionation (or backward reactions) via
ion-molecule reactions would proceed within the disk depending on the
local ionization rate, temperature, and degree of molecular depletion,
while the vertical and radial diffusion will tend to lessen the
spatial gradient of D/H ratios.  Inclusion of deuterated species in
the recent models with non-thermal chemistry and 2-D diffusion is
desirable.

\section{\textbf{Outstanding Issues and Future Prospectus}}

Significant gains have been made in our observational and theoretical
understanding of the chemical evolution of protoplanetary disks in the
decade since the last review in this conference series.  We now
recognize the importance of the irradiated surface, which at the very
least contains an active chemistry and is responsible for most
observed molecular emission lines.  The gross characteristics and key
ingredients of this surface are roughly understood.  How the varied
effects (grain evolution, UV/X-ray radiation dominance, etc..) play
out on the chemical evolution in terms of X-ray/UV dominance and the
dependence on other evolutionary factors, is one of the challenges for
future models.
%
Given the observed chemical complexity, a detailed
understanding of the chemistry is a pre-requisite for the interpretation
of ongoing and future observations of molecular emission in
protoplanetary disks.

Disk surface processes may dominate the observed chemistry, 
but it is not certain how much of a role this chemistry plays in altering the
chemical characteristics within the primary mass reservoir, the disk midplane.
Thus one of the main outstanding questions for disk chemistry remains:
how much material remains pristine and chemically unaltered from its origin
in the parent molecular cloud.
We now have observational and theoretical evidence for active chemical
zones; thus it is likely that the most volatile species, 
which are frozen on grains
in the infalling material (e.g. CO, N$_2$) do undergo
significant processing.  This will trickle down to other, less
abundant molecules that
form easily from the ``parent species'' (e.g. H$_2$CO, HCN and the
deuterated counterparts).  
Disentangling these effects will be 
complicated because the chemistry in the outer
disk ($r >$ 30 AU), which through advection feeds the inner disk,  is
quite similar to that seen in dense regions of the interstellar medium.
For the least volatile
molecules, in particular water ice, sublimation and subsequent gas-phase
alteration is less likely, unless there is significant radial mixing
from the warmer inner disk to colder outer regions.

In part, our recognition of the warm molecular layer and the importance
of photo-chemistry, is driven by our
current observational facilities, which are unable to resolve the
innermost regions of the disk (e.g. planet forming zones), coupling
better to the larger surface area of the outer disk.
Within $r <$ 10--30 AU,  the
midplane and surface are both hot enough to sublime even the least
volatile molecules (e.g. H$_2$O), eventually
producing an active chemistry that is 
described by the earlier PP III review ({\it Prinn},
1993) and in the Chapter by Najita et al.
The transition between these layers, the so-called ``snow-line'' and the
chemistry within the planet forming zone, will
be species specific and should
be readily detectable with upcoming advances in our capabilities, in
particular the eagerly awaited ALMA array.  

In summary, we stand on the cusp of the marriage of a rapidly emerging
new field, studies
of extra-solar protoplanetary disk chemical evolution, and an old one,
the cosmochemical study of planets, meteorites, asteroids, and comets.
In this review we have outlined broad areas where the evolving chemistry
can be altered through changes induced by vertical and
horizontal temperature gradients, the evolution of grain properties, 
and disk dynamics (mixing).  
Thus studies of active chemistry in extra-solar 
disks offers the
promise and possibility to untangle long-standing questions regarding 
the initial conditions, chemistry, and dynamics of planet formation, 
the origin of cometary ices, and, ultimately, a greater understanding of
the organic content of gas/solid reservoirs that produced life  at least
once in the Galaxy.

\acknowledgements

E.A.B. thanks L. Hartmann for an initial reading.  We also acknowledge
gratefully receipt of unpublished material from B. Jonkheid, C. Qi,
D. Semenov, N. Turner, and K. Willacy.  This work was supported by in
part by NASA through grants NNG04GH27G and 09374.01-A from STScI, by a
Grant-in-Aid for Scientific Research (17039008) and ``The 21st Century
COE Program of the Origin and Evolution of Planetary Systems'' of the
Ministry of Education, Culture, Sports, Science, and Technology (MEXT)
of Japan, and by a Spinoza award of NWO.

\centerline\textbf{ REFERENCES}
\bigskip
\parskip=0pt
{\small
\baselineskip=11pt

\refs Acke, B., van den Ancker,
M.~E., Dullemond, C.~P.\ 2005.\ 
{\em Astron. Astrophys.,  436},
209-230.

\refs Acke, B., van
den Ancker, M.~E.\ 2004.\ 
{\em Astron. Astrophys., 426}, 151-170.


\refs Aikawa, Y., Nomura, H. 2006, ApJ, in press


\refs Aikawa, Y., van
Zadelhoff, G.~J., van Dishoeck, E.~F., Herbst, E.\ 2002.\ 
{\em Astron. Astrophys., 386}, 622-632.

\refs Aikawa, Y., Herbst,
E.\ 2001.\ 
{\em Astron. Astrophys., 371}, 1107-1117.

\refs Aikawa, Y., Herbst,
E.\ 1999.\ 
{\em
Astron. Astrophys.,  351}, 233-246.

\refs Aikawa, Y., Umebayashi,
T., Nakano, T., Miyama, S.~M.\ 1999.\ 
{\em Astrophys. J., 519},
705-725.

\refs Aikawa, Y., Umebayashi,
T., Nakano, T., Miyama, S.~M.\ 1997.\ 
{\em Astrophys. J., 486}, L51.

\refs Alexander, R.~D.,
Clarke, C.~J., Pringle, J.~E.\ 2005.\ 
{\em Mon. Not. Roy. Astron.
Soc., 358}, 283-290.

\refs Bacmann, A., Lefloch,
B., Ceccarelli, C., Steinacker et. al. \ 2003.\ CO
{\em Astrophys.  J., 585}, L55-L58.

\refs Balbus, S.~A.,
Papaloizou, J.~C.~B.\ 1999.\ 
{\em Astrophys. J., 521}, 650-658.

\refs Bally, J., Sutherland,
R.~S., Devine, D., Johnstone, D.\ 1998.\ 
{\em Astron. J., 116}, 293-321.


\refs Bary, J.~S., Weintraub, 
D.~A., Kastner, J.~H.\ 2003.\ 
{\em Astrophys. J., 586}, 1136-1147.

\refs Beckwith, S.~V.~W.,
Henning, T., Nakagawa, Y.\ 2000.\ 
In {\em Protostars and Planets IV}
(V. Mannings, A. P. Boss, and S. S. Russell, eds.),
Univ. of Arizona Press, Tucson, 533.


\refs Beckwith, S.~V.~W.,
Sargent, A.~I., Chini, R.~S., Guesten, R.\ 1990.\ 
{\em Astron. J., 99},
924-945.

\refs Bergin, E., and 11
colleagues 2004.\ 
{\em Astrophys. J., 614}, L133-L136.

\refs Bergin, E., Calvet, N.,
D'Alessio, P., Herczeg, G.~J.\ 2003.\ 
{\em Astrophys. J., 591}, L159-L162.

\refs Blake, G.~A.,
Boogert, A.~C.~A.\ 2004.\ 
{\em Astrophys. J., 606}, L73-L76.

\refs Blake, G.~A., Qi, C.,
Hogerheijde, M.~R., Gurwell, M.~A., Muhleman, D.~O.\ 1999.\ 
{\em Nature, 398}, 213.

\refs
Bockel{\'e}e-Morvan, D., Gautier, D., Hersant, F., Hur{\'e}, J.-M., Robert,
F.\ 2002.\ {\em Astron. Astrophys., 384},
1107-1118.

\refs
Bockel{\'e}e-Morvan, D., and 11 colleagues 1998.\ 
{\em Icarus, 133}, 147-162.

\refs Boogert, A.~C.~A.,
Hogerheijde, M.~R., Blake, G.~A.\ 2002.\ 
{\em Astrophys. J., 568}, 761-770.

\refs Bringa, E.~M.,
Johnson, R.~E.\ 2004.\ 
{\em Astrophys. J., 603},
159-164.

\refs Brittain, S.~D.,
Rettig, T.~W.\ 2002.\ 
{\em Nature, 418}, 57-59.

\refs Calvet, N., D'Alessio,
P., Hartmann, L., Wilner, D., Walsh, A., Sitko, M.\ 2002.\ 
{\em Astrophys. J., 568}, 1008-1016.

\refs Calvet, N.,
Gullbring, E.\ 1998.\ 
{\em Astrophys. J., 509}, 802-818.

\refs Calvet, N., Magris,
G.~C., Patino, A., D'Alessio, P.\ 1992.\ 
{\em Revista Mexicana
de Astronomia y Astrofisica, 24}, 27.

\refs Carballido, A.,
Stone, J.~M., Pringle, J.~E.\ 2005.\ 
{\em Mon. Not. Roy. Astr. Soc., 358}, 1055-1060.

\refs Carr, J.~S., Tokunaga,
A.~T., Najita, J.\ 2004.\ 
{\em Astrophys. J., 603}, 213-220.

\refs Caselli, P.\ 2002.\ 
{\em Planetary and Space Science, 50}, 1133-1144.

\refs Caselli, P., Walmsley,
C.~M., Zucconi, A., Tafalla, M., Dore, L., Myers, P.~C.\ 2002.\ 
{\em Astrophys. J., 565},
344-358.

\refs Ceccarelli, C.,
Dominik, C., Caux, E., Lefloch, B., Caselli, P.\ 2005.\ 
{\em Astrophys. J., 631}, L81-L84.

\refs Ceccarelli, C.,
Dominik, C.\ 2005.\ 
{\em Astron. Astrophys., 440}, 583-593.

\refs Ceccarelli, C.,
Dominik, C., Lefloch, B., Caselli, P., Caux, E.\ 2004.\ 
 {\em Astrophys. J., 607}, L51-L54.

\refs Chiang, E.~I., Joung,
M.~K., Creech-Eakman, M.~J., Qi, C., Kessler, J.~E., Blake, G.~A., van
Dishoeck, E.~F.\ 2001.\ 
{\em Astrophys.
J., 547}, 1077-1089.

\refs Chiang, E.~I.,
Goldreich, P.\ 1997.\ 
{\em Astrophys. J., 490}, 368.

\refs Crovisier, J.,
Bockel{\'e}e-Morvan, D., Colom, P., Biver, N., Despois, D., Lis, D.~C. 2004.
{\em Astron. Astrophys., 418}, 1141-1157.

\refs Crovisier, J., Leech,
K., Bockel{\'e}e-Morvan, D., Brooke, T. et. al. \ 1997.\ 
{\em Science, 275}, 1904-1907.


\refs D'Alessio, P.,
Calvet, N., Hartmann, L., Lizano, S., Cant{\'o}, J.\ 1999.\ 
{\em Astrophys. J., 527}, 893-909.

\refs D'Alessio, P., Canto,
J., Calvet, N., Lizano, S.\ 1998.\ 
{\em Astrophys. J., 500}, 411.

\refs Dartois, E., A.~Dutrey, and
S.~Guilloteau 2003.\ 
{\em Astron. Astrophys., 399}, 773-787.

\refs Draine, B.~T., Sutin,
B.\ 1987.\ {\em Astrophys. J., 320}, 803-817.

\refs Draine, B. T. 1978, 
{\it Astrophys. J. Suppl., 36}, 595-619


\refs Dullemond,
C.~P., Dominik, C.\ 2005.\ 
{\em Astron. Astrophys., 434}, 971-986.

\refs Dullemond,
C.~P., Dominik, C.\ 2004.\ 
{\em Astron. Astrophys., 421}, 1075-1086.

\refs Dupree, A.~K.,
Brickhouse, N.~S., Smith, G.~H., Strader, J.\ 2005.\ 
{\em Astrophys. J., 625},
L131-L134.

\refs Duschl, W.~J., Gail,
H.-P., Tscharnuter, W.~M.\ 1996.\ 
{\em Astron. Astrophys., 312}, 624-642.

\refs Dutrey, A., Guilloteau, S.\ 2004. {\em Astrophys. Space Sci.,
292}, 407-418.

\refs
Dutrey, A., Guilloteau, S., Simon, M.\ 1994.\ {\em Astron. Astrophys., 291},
L23-L26.

\refs Dutrey, A., Guilloteau,
S., Guelin, M.\ 1997.\ 
{\em Astron. Astrophys., 317},
L55-L58.

\refs Duvert, G. et al.\ 2000.\ {\em Astron. Astrophys., 355}, 165-170.
 
\refs Eberhardt, P., Reber,
M., Krankowsky, D., Hodges, R.~R.\ 1995.\ 
{\em Astron.
Astrophys., 302}, 301.

\refs Favata, F., Flaccomio,
E., Reale, F., Micela, G., Sciortino, S. et. al. \ 2005.\ 
{\em Astrophys. J.,
Supp. Ser., 160}, 469-502.

\refs Fegley, B.~J.\ 1999.\ 
{\em Space Sci. Rev., 90}, 239-252.

\refs Fegley, B.~J., Prinn,
R.~G.\ 1989.\ In
 {\em The Formation and Evolution of Planetary Systems}, 171-205.

\refs Feigelson,
E.~D., Montmerle, T.\ 1999.\ 
{\em Ann. Rev. Astron. Astrophys., 37}, 363-408.

\refs Finocchi, F., Gail,
H.-P.\ 1997.\ 
{\em Astron. Astrophys., 327},
825-844.

\refs Forrest, W.~J., and 20
colleagues 2004.\ 
{\em Astrophys. J. Suppl. Ser., 154}, 443-447.

\refs Gammie, C.~F.\ 2001.\ 
{\em Astrophys. J., 553}, 174-183.

\refs Gammie, C.~F.\ 1996.\ 
{\em Astrophys. J., 457}, 355.

\refs Gensheimer, P.~D.,
Mauersberger, R., Wilson, T.~L.\ 1996.\ 
{\em Astron. Astrophys., 314}, 281-294.


\refs Glassgold, A.~E., Feigelson, E.D., Montmerle, T. 2000,
In {\em Protostars and Planets IV},
(V. Mannings, A. P. Boss, and S. S. Russell, eds.),
Univ. of Arizona Press, Tucson, 429.

\refs Glassgold, A.~E.,
Najita, J., Igea, J.\ 2004.\ 
{\em Astrophys. J., 615}, 972-990.

\refs Glassgold, A.~E.,
Najita, J., Igea, J.\ 1997.\ 
{\em Astrophys. J., 480}, 344.

\refs Goldsmith, P.~F.\ 2001.\
{\em Astrophys. J., 557}, 736-746.


\refs Goto, M., Geballe,
T.~R., 
McCall, B.~J., Usuda, T. et. al. 
2005.\ {\em Astrophys. J., 629}, 
865-872. 
 
\refs Habart, E., Natta, A.,
Kr{\"u}gel, E.\ 2004.\ 
{\em Astron. Astrophys., 427}, 179-192.

\refs Habing, H.~J.\ 1968.\ 
{\em Bull. Astron. Inst. Neth., 19}, 421.

\refs Hartmann, L., Calvet,
N., Gullbring, E., D'Alessio, P.\ 1998.\ 
{\em Astrophys. J., 495}, 385.

\refs Hasegawa, T.~I.,
Herbst, E.\ 1993.\ 
{\em Mon. Not. Roy. Astro. Soc., 261},
83-102.

\refs Hatchell, J.,
Thompson, M.~A., Millar, T.~J., MacDonald, G.~H.\ 1998.\ 
{\em Astron.
Astrophys. Supp. Ser., 133}, 29-49.

\refs Hawley, J.~F.,
Balbus, S.~A.\ 1991.\ 
{\em Astrophys. J., 376}, 223.

\refs Hayashi, C.\ 1981.\ 
{\em Prog. Theor.
Phys. Suppl., 70}, 35-53.

\refs Herbig, G.~H.,
Goodrich, R.~W.\ 1986.\ 
{\em Astrophys. J., 309}, 294-305.

\refs Herbst, E.,
Klemperer, W.\ 1973.\ 
{\em Astrophys. J., 185}, 505-534.

\refs Herczeg, G.~J., Wood,
B.~E., Linsky, J.~L., Valenti, J.~A., Johns-Krull, C.~M.\ 2004.\ 
{\em Astrophys. J., 607}, 369-383.

\refs Hersant, F., Gautier,
D., Hur{\'e}, J.-M.\ 2001.\ 
{\em Astrophys. J., 554}, 391-407.

\refs Hester, J.J., and Desch, S.J. 2006, In
{\em Proceedings from the Workshop on Chondrites and the Protoplanetary
Disk}, (A. Krot, E. Scott, and B. Reipurthi, eds.), ASP Conference
Series, in press.

\refs Hollenbach, D.~J.,
Yorke, H.~W., Johnstone, D.\ 2000.\ 
In {\em Protostars and Planets IV},
(V. Mannings, A. P. Boss, and S. S. Russell, eds.),
Univ. of Arizona Press, Tucson, 401.

\refs Hollenbach,
D.~J., Tielens, A.~G.~G.~M.\ 1999.\ 
{\em Rev. Mod. Phys., 71}, 173-230.

\refs Igea, J.,
Glassgold, A.~E.\ 1999.\ 
{\em Astrophys. J., 518}, 848-858.

\refs Ilgner, M.~and
R.~P.~Nelson 2005b.\ 
{\em Astron. Astrophys.},
in press.

\refs
Ilgner, M.~and
R.~P.~Nelson 2005a.\ 
{\em Astron. Astrophys.},
in press.

\refs Ilgner, M., Henning, T.,
Markwick, A.~J., Millar, T.~J.\ 2004.\ 
{\em Astron. Astrophys., 415},
643-659.

\refs Inutsuka, S.-i.~and
T.~Sano 2005.\ 
{\em Astrophys. J., 628}, L155-L158.

\refs Irvine, W.~M.\ 1999.\ 
{\em Space Sci. Rev., 90},
203-218.

%

\refs Jonkheid, B., Faas,
F.~G.~A., van Zadelhoff, G.-J., van Dishoeck, E.~F.\ 2004.\ 
{\em Astron. Astrophys., 428}, 511-521.

\refs Jonkheid, B., Kamp, I., Augereau, J.-C., van Dishoeck, E.~F.\ 2006.\
{\em Astron. Astrophys.,} submitted.

\refs J{\o}rgensen, 
J.~K., Sch{\"o}ier, F.~L., and van Dishoeck, E.~F.\ 2004, {\em Astron. Astrophys.}, 416, 603

\refs Kamp, I.,
Dullemond, C.~P.\ 2004.\ 
{\em Astrophys. J., 615}, 991-999.

\refs Kamp, I., van
Zadelhoff, G.-J.\ 2001.\ 
{\em Astron. Astrophys., 373}, 641-656.

\refs Kastner, J.~H., Franz,
G., Grosso, N., Bally, J. et. al. \ 2005.\ 
{\em Astrophys. J. Suppl. Ser.,
160}, 511-529.

\refs Kastner, J.~H.,
Zuckerman, B., Weintraub, D.~A., Forveille, T.\ 1997.\ 
{\em Science, 277},
67-71.

\refs Kenyon, S.~J.,
Hartmann, L.\ 1987.\ 
{\em Astrophys. J., 323}, 714-733.

\refs
Kessler-Silacci, J.~E., et al.\ 2006. 
{\em Astrophys. J., 639}, 275-291.

\refs
Kessler-Silacci, J.~E., Hillenbrand, L.~A., Blake, G.~A., Meyer, M.~R.\
2005.\ 
{\em Astrophys. J., 622}, 404-429.

\refs Kessler, J.~E., Blake,
G.~A., Qi, C.\ 2003.\ 
in {\em Chemistry as a Diagnostic of Star Formation}
(C. L.~Curry and M.Fich, eds), ~ NRC
Press, Ottawa, Canada, 2003, p.~188.\ 188.

\refs Kitamura, Y., M.~Momose,
S.~Yokogawa, R.~Kawabe, M.~Tamura, and S.~Ida 2002.\ 
{\em Astrophys. J., 581}, 357-380.

\refs Klahr, H.~H.,
Bodenheimer, P.\ 2003.\ 
{\em Astrophys. J., 582}, 869-892.

\refs Koerner, D.M., Sargent, A.I., Beckwith, S.V.W.\ 1993.\ {\em Icarus,
106}, 2-12.

\refs Lada, C.~J., Lada, 
E.~A.\ 2003.\ {\em Ann. Rev.
Astron. Astrophys., 41}, 57-115.

\refs Lahuis, F., et al.\ 2006.\ {\em Astrophys. J., 636}, L145-148.

\refs Larsson, M., Danared,
H., Larson, {\AA}., Le Padellec, A., Peterson, J.R. et. al.\ 1997.\ 
{\em Phys. Rev. Let., 79},
395-398.

\refs L\'eger, A., Jura, M.,
Omont, A.\ 1985.\ 
{\em Astron. Astrophys., 144}, 147-160.

\refs Le Teuff, Y.~H.,
Millar, T.~J., Markwick, A.~J.\ 2000.\ 
{\em Astron. Astrophys. Suppl. Ser., 146},
157-168.

\refs Lewis, J.~S., Prinn,
R.~G.\ 1980.\ 
{\em Astrophys. J., 238}, 357-364.

\refs Li, A., Lunine, J.~I.\
2003.\ 
{\em Astrophys. J., 594}, 987-1010.

\refs MacPherson, G.~J.,
Wark, D.~A., Armstrong, J.~T.\ 1988.\ 
In {\em Meteorites and the Early Solar System},
746-807.

\refs Malfait, K., Waelkens,
C., Waters, L.~B.~F.~M., Vandenbussche, B. et. al. 1998.\ 
{\em Astron. Astrophys., 332}, L25-L28.

\refs Maloney, P.~R.,
Hollenbach, D.~J., Tielens, A.~G.~G.~M.\ 1996.\ 
{\em Astrophys. J., 466},
561.

\refs Markwick, A.~J.,
Ilgner, M., Millar, T.~J., Henning, T.\ 2002.\ 
{\em Astron. Astrophys., 385},
632-646.

\refs McCall, B.~J., and 15
colleagues 2004.\ 
{\em Phys. Rev. A, 70}, 052716.

\refs Meier, R., Owen, T.~C.,
Jewitt, D.~C., Matthews, H.~E., Senay, M., Biver, N., Bockel{\'e}e-Morvan, D.,
Crovisier, J., Gautier, D.\ 1998.\ 
{\em Science, 279}, 1707.

\refs Millar, T.~J., Bennett,
A., Herbst, E.\ 1989.\ 
 {\em Astrophys. J., 340}, 906-920.


\refs Morrison, R.,
McCammon, D.\ 1983.\ 
{\em Astrophys. J., 270}, 119-122.

\refs Mumma, M.~J., Weissman,
P.~R., Stern, S.~A.\ 1993.\ 
In {\em Protostars and Planets III},
(E. Levy and J. Lunine, eds.),
Univ. of Arizona Press, Tucson, 1177-1252.

\refs Najita, J., Carr, J.~S.,
Mathieu, R.~D.\ 2003.\ 
{\em Astrophys. J., 589},
931-952.

\refs Najita, J., Bergin,
E.~A., Ullom, J.~N.\ 2001.\ 
{\em Astrophys. J., 561}, 880-889.

\refs Nakagawa, Y.,
Nakazawa, K., Hayashi, C.\ 1981.\ 
{\em Icarus, 45}, 517-528.

\refs Nishi, R., Nakano, T.,
Umebayashi, T.\ 1991.\ 
{\em Astrophys. J., 368}, 181-194.

\refs Nomura, H.\ 2002.\ 
{\em Astrophys. J., 567}, 587-595.

\refs Nomura, H.~and
T.~J.~Millar 2005.\ 
{\em Astron. Astrophys, 438}, 923-938.

\refs Parise, B., and 14
colleagues 2005.\ 
{\em Astron. Astrophys., 431}, 547-554.

\refs Parise, B., Castets, A.,
Herbst, E., Caux, E., Ceccarelli, C. et. al. \ 2004.\ 
{\em Astron. Astrophys., 416}, 159-163.

\refs Parise, B., Ceccarelli,
C., Tielens, A.~G.~G.~M., Herbst, E. et. al. \ 2002.\ 
{\em Astron. Astrophys., 393}, L49-L53.

\refs Pontoppidan, K.~M.,
Dullemond, C.~P., van Dishoeck, E.~F., Blake, G.~A.
et. al.  2005.\ 
{\em Astrophys. J., 622}, 463-481.

\refs Pringle, J.~E.\ 1981.\
{\em Ann. Rev. Astron.  Astrophy., 19}, 137-162.

\refs Prinn, R.~G.\ 1993.\ 
In {\em Protostars and Planets III},
(E. Levy and J. Lunine, eds.),
Univ. of Arizona Press, Tucson,
1005-1028.

\refs Prinn, R.~G.\ 1990.\ 
{\em Astrophys. J., 348}, 725-729.

\refs Prinn, R.~G., Fegley,
B.\ 1981.\ 
{\em Astrophys. J.,
249}, 308-317.

\refs Przygodda, F., van
Boekel, R., {\`A}brah{\`a}m, P., Melnikov, S.~Y. et. al. \ 2003.\ 
{\em Astron. Astrophys., 412}, L43-L46.

\refs Qi., C. et al.\ 2004.\ {\em Astrophys. J., 616}, L7-10.

\refs Qi, C., Kessler, J.~E.,
Koerner, D.~W., Sargent, A.~I., Blake, G.~A.\ 2003.\ 
{\em Astrophys. J., 597}, 986-997.

\refs Richter, M.~J., Jaffe,
D.~T., Blake, G.~A., Lacy, J.~H.\ 2002.\ 
{\em Astrophys. J., 572}, L161-L164.

\refs Roberts, H., Herbst,
E., Millar, T.~J.\ 2004.\ 
{\em Astron. Astrophys., 424}, 905-917.

\refs Roberts, H., Fuller,
G.~A., Millar, T.~J., Hatchell, J., Buckle, J.~V.\ 2002.\ 
{\em Astron. Astrophys., 381}, 1026-1038.

\refs
Rodriguez-Franco, A., Mart{\'i}n-Pintado, J., Fuente, A.\ 1998.\ 
{\em Astron. Astrophys., 329}, 1097-1110.

\refs Roueff, E., Lis, D.~C.,
van der Tak, F.~F.~S., Gerin, M., Goldsmith, P.~F.\ 2005.\ 
{\em Astron. Astrophys., 438},
585-598.

\refs Sako, S., Yamashita, T.,
Kataza, H., Miyata, T., Okamoto, Y.~K., Honda, M., Fujiyoshi, T., Onaka,
T.\ 2005.\ Search for 17 {$\mu$}m H$_{2}$ 
{\em Astrophys. J., 620}, 347-354.

\refs Sano, T., Miyama, S.~M.,
Umebayashi, T., Nakano, T.\ 2000.\ 
{\em Astrophys. J., 543}, 486-501.

\refs Schutte, W.~A.,
Khanna, R.~K.\ 2003.\ 
{\em Astron. Astrophys., 398}, 1049-1062.

\refs Semenov, D., Wiebe, D.,
Henning, T.\ 2004.\ 
{\em Astron. Astrophys.,
417}, 93-106.

\refs Shakura, N.~I.,
Sunyaev, R.~A.\ 1973.\ 
{\em Astron. Astrophys., 24}, 337-355.

\refs Shen, C.~J., Greenberg, J.~M., Schutte, W.~A., van Dishoeck, E.~F.\
2004. {\em Astron. Astrophs., 415}, 203-215.

\refs Simon, M., Dutrey, A.,
Guilloteau, S.\ 2000.\ 
{\em Astrophys. J., 545}, 1034-1043.

\refs St{\"a}uber, P.,
Doty, S.~D., van Dishoeck, E.~F., Benz, A.~O.\ 2005.\ 
{\em Astron. Astrophys.,
440}, 949-966.

\refs Stevenson, D.~J.\ 1990.\
{\em Astrophys. J., 348}, 730-737.

\refs Stone, J.~M., Gammie,
C.~F., Balbus, S.~A., Hawley, J.~F.\ 2000.\ 
In {\em Protostars and Planets IV},
(V. Mannings, A. P. Boss, and S. S. Russell, eds.),
Univ. of Arizona Press, Tucson, 589.

\refs St{\"o}rzer, H.,
Hollenbach, D.\ 1998.\ 
{\em Astrophys.  J., 502}, L71.

\refs Thi, W.-F., van Zadelhoff,
G.-J., van Dishoeck, E.~F.\ 2004.\ 
{\em Astron. Astrophys., 425},
955-972.

\refs Thi, W.~F., Pontoppidan,
K.~M., van Dishoeck, E.~F., Dartois, E., d'Hendecourt, L.\ 2002.\ 
{\em Astron. 
Astrophys., 394}, L27-L30.

\refs Thi, W.~F., and 10
colleagues 2001.\ 
{\em Astrophys. J., 561}, 1074-1094.

\refs Tin{\'e}, S., Roueff,
E., Falgarone, E., Gerin, M., Pineau des For{\^e}ts, G.\ 2000.\ 
{\em Astron. Astrophys., 356},
1039-1049.

\refs Turner, N.J., Willacy, K., Bryden, G., and Yorke, H.W.
\ 2006.\ 
{\em Astrophys. J.}, in press.

\refs Uchida, K.~I., and 19
colleagues 2004.\ 
{\em Astrophys. J. Suppl. Ser., 154}, 439-442.

\refs Umebayashi, T.,
Nakano, T.\ 1981.\ 
{\em Pub. Astron. Soc. Japan, 33}, 617.

\refs van Boekel, R., Min, M.,
Waters, L.~B.~F.~M., de Koter, A., Dominik, C., van den Ancker, M.~E.,
Bouwman, J.\ 2005.\ 
{\em Astron. Astrophys., 437}, 189-208.

\refs van Dishoeck, E.~F.\
2004.\ 
{\em Ann. Rev. Astron. Astrophys., 42},
119-167.

\refs van Dishoeck,
E.~F., Thi, W.-F., van Zadelhoff, G.-J.\ 2003.\ 
{\em Astron. Astrophys., 400}, L1-L4.

\refs van Dishoeck, E.~F.,
Blake, G.~A., Draine, B.~T., Lunine, J.~I.\ 1993.\ 
In {\em Protostars and Planets III},
(E. Levy and J. Lunine, eds.),
Univ. of Arizona Press, Tucson,
163-241.

\refs van
Dishoeck, E.~F., Dalgarno, A.\ 1984.\ 
{\em Icarus, 59}, 305-313.

\refs van Dishoeck, E.~F., Jonkheid, B., van Hemert, M.C.\ 2006.\
{\em Faraday Discussions, 133}, in press.

\refs van Zadelhoff, G.J., van Dishoeck, E.F., Thi, W.F., Blake, G.A.\
2001.\ {\em Astron. Astrophys., 377}, 566-580.

\refs van Zadelhoff,
G.-J., Aikawa, Y., Hogerheijde, M.~R., van Dishoeck, E.~F.\ 2003.\
{\em Astron. Astrophys., 397}, 789-802.

\refs Walmsley, C.~M.,
Flower, D.~R., Pineau des For{\^e}ts, G.\ 2004.\ 
{\em Astron. Astrophys., 418}, 1035-1043.

\refs Watson, D.~M., and 20
colleagues 2004.\ 
{\em Astrophys. J. Suppl. Ser., 154}, 391-395.

\refs Weidenschilling,
S.~J.\ 1997.\ 
{\em Icarus, 127}, 290-306.

\refs
Weidenschilling, S.~J., Cuzzi, J.~N.\ 1993.\ 
In {\em Protostars and Planets III},
(E. Levy and J. Lunine, eds.),
Univ. of Arizona Press, Tucson, 1031-1060.

\refs Weingartner,
J.~C., Draine, B.~T.\ 2001.\ 
{\em Astrophys. J., 563}, 842-852.

\refs Wehrstedt, M.,
Gail, H.-P.\ 2003.\ 
{\em Astron. Astrophys., 410},
917-935.

\refs Westley, M.~S.,
Baragiola, R.~A., Johnson, R.~E., Baratta, G.~A.\ 1995.\ 
{\em Nature, 373}, 405.

\refs Willacy, K.,
Langer, W.~D., and Allen, M.\ 2006.\ 
{\em Astrophys. J.}, in press.

\refs Willacy, K.,
Langer, W.~D.\ 2000.\ 
{\em Astrophys. J., 544}, 903-920.

\refs Willacy, K., Klahr,
H.~H., Millar, T.~J., Henning, T.\ 1998.\ 
{\em Astron. Astrophys., 338}, 995-1005.


\end{document}